\documentclass[10pt,journal]{IEEEtran}

\ifCLASSOPTIONcompsoc
  \usepackage[compress]{cite}
\else
  \usepackage{cite}
\fi

\usepackage{ragged2e}
\usepackage{underscore}
\usepackage{pifont}
\usepackage{adjustbox}
\usepackage{hyperref}
\usepackage{amssymb}
\usepackage{graphicx}
\usepackage{multirow}
\usepackage{bbding}
\usepackage{float}
\usepackage{empheq}
\usepackage{amsmath}
\usepackage{cancel}
\usepackage{url}
\usepackage{graphicx}
\usepackage{color}
\usepackage[vlined,ruled]{algorithm2e}
\SetAlgoNlRelativeSize{0} 

\usepackage{setspace}
\usepackage{stfloats}
\usepackage{multirow}
\usepackage{makecell}
\usepackage{multicol}
\usepackage[noend]{algpseudocode}
\SetAlgoNoLine

\algdef{SE}[PROCEDURE]{Procedure}{EndProcedure}[2]{\textbf{procedure} #1(#2)}{\textbf{end procedure}}

\usepackage{bm}
\usepackage{float}
\usepackage[table,xcdraw]{xcolor}
\usepackage[normalem]{ulem}

\usepackage{amsmath}

\everymath{\displaystyle}
\everydisplay{\displaystyle}

\let\oldsum\sum
\renewcommand{\sum}{\displaystyle\oldsum}

\useunder{\uline}{\ul}{}

\usepackage{float}  
\usepackage{lipsum}


\ifCLASSINFOpdf
\else
\fi
\begin{document}
%

\title{Recursive Offloading for LLM Serving in Multi-tier Networks}
%
%

\author{Zhiyuan~Wu,
        Sheng~Sun,
        Yuwei~Wang,~\IEEEmembership{Member,~IEEE,}
        Min~Liu,~\IEEEmembership{Senior~Member,~IEEE,}\\
        Bo~Gao,~\IEEEmembership{Member,~IEEE,}
        Jinda~Lu,
        Zheming~Yang,
        and~Tian~Wen
\IEEEcompsocitemizethanks{
\IEEEcompsocthanksitem Zhiyuan Wu and Tian Wen are with the State Key Lab of Processers, Institute of Computing Technology, Chinese Academy of Sciences, Beijing, China, and also with the University of Chinese Academy of Sciences, Beijing, China.
E-mails: \{wuzhiyuan22s, wentian24s\}@ict.ac.cn. 
\IEEEcompsocthanksitem Sheng Sun, Yuwei Wang and Mingzhe Yang are with the State Key Lab of Processers, Institute of Computing Technology, Chinese Academy of Sciences, Beijing, China.
E-mails: \{sunsheng, ywwang, yangzheming\}@ict.ac.cn.
\IEEEcompsocthanksitem Min Liu is with the State Key Lab of Processers, Institute of Computing Technology, Chinese Academy of Sciences, Beijing, China, and also with the Zhongguancun Laboratory, Beijing, China.
E-mail: liumin@ict.ac.cn
\IEEEcompsocthanksitem Bo Gao is with the School of Computer Science and Technology, and the Engineering Research Center of Network Management Technology for High-Speed Railway of Ministry of Education, Beijing Jiaotong University, Beijing, China.
E-mail: bogao@bjtu.edu.cn.
\IEEEcompsocthanksitem Jinda Lu is with the School of Cyber Science and Technology, University of Science and Technology of China, Hefei, China.
E-mail: lujd@mail.ustc.edu.cn.
\IEEEcompsocthanksitem Corresponding author: Yuwei Wang.
}
\thanks{This work was supported by the National Key Research and Development Program of China (No. 2023YFB2703701), the National Natural Science Foundation of China (No. 62472410).}
}

%
%

\markboth{Under Review}%
{Shell \MakeLowercase{\textit{et al.}}: Bare Demo of IEEEtran.cls for Computer Society Journals}
%




\IEEEtitleabstractindextext{%
\begin{abstract}
\justifying
Heterogeneous device-edge-cloud computing infrastructures have become widely adopted in telecommunication operators and Wide Area Networks (WANs), providing multi-tier computational support for emerging intelligent services. With the rapid proliferation of Large Language Model (LLM) services, efficiently coordinating inference tasks and reducing communication overhead within these multi-tier network architectures becomes a critical deployment challenge. 
\textcolor{black}{Current LLM serving paradigms exhibit significant limitations: on-device deployment is constrained to lightweight LLMs by hardware capabilities, while cloud-centric deployment encounters resource congestion and considerable prompt communication overhead during peak periods.}
\textcolor{black}{Model-cascading inference, though better suited to multi-tier networks, depends on static, manually-tuned thresholds that cannot adapt to dynamic network conditions or varying task complexities.}
To address these challenges, we propose RecServe, a recursive offloading framework tailored for LLM serving in multi-tier networks.
\textcolor{black}{RecServe introduces a task-specific hierarchical confidence evaluation mechanism that guides offloading decisions based on inferred task complexity in progressively scaled LLMs across device, edge, and cloud tiers.} To further enable intelligent task routing across tiers, RecServe employs a sliding-window-based dynamic offloading strategy with quantile interpolation, enabling real-time tracking of historical confidence distributions and adaptive offloading threshold adjustments. \textcolor{black}{This design allows inference tasks to be recursively offloaded to higher tiers only when necessary, optimizing heterogeneous resource utilization while reducing cross-tier communication with little compromise on service quality.}
Theoretical analysis provides distinct conditions under which RecServe is expected to achieve reduced communication burden and computational costs.
Experiments on eight datasets demonstrate that RecServe outperforms CasServe in both service quality and communication efficiency, and reduces the communication burden by over 50\% compared to centralized cloud-based serving. Our code is available at \url{https://github.com/wuzhiyuan2000/RecServe}.

\end{abstract}

\begin{IEEEkeywords}
Large language models, services computing, task offloading, edge-cloud collaboration, communication efficiency
\end{IEEEkeywords}
}
\maketitle

\IEEEdisplaynontitleabstractindextext

%

\IEEEpeerreviewmaketitle
\vspace{20pt}
\IEEEraisesectionheading{\section{Introduction}}
\IEEEPARstart{T}{he} rapid proliferation of the Internet of Things (IoT), Mobile Edge Computing (MEC), and Artificial Intelligence (AI) has driven unprecedented demand for intelligent services. To meet the computational requirements of these applications, telecommunication operators and Wide Area Networks (WANs) increasingly adopt multi-tier heterogeneous computing infrastructures consisting of end devices, edge computing nodes, and cloud data centers \cite{ren2019survey}. 
This hierarchical architecture exhibits an inverse relationship between computational capacity and user proximity: end devices directly interact with user data but typically possess limited computing power; edge nodes offer moderate processing capabilities with low-latency connectivity; cloud data centers host powerful CPU/GPU clusters but are situated remotely from data sources \cite{duan2022distributed,wang2024end}. Such multi-tier network systems offer remarkable flexibility for optimizing resource allocation, minimizing service latency, and enhancing user experience.

Among the emerging intelligent applications, services powered by Large Language Models (LLMs) have demonstrated remarkable impacts. Built upon Transformer \cite{radford2019language} architecture and self-attention \cite{vaswani2017attention} mechanism, LLMs effectively capture long-range contextual dependencies within text, achieving capabilities that match or exceed human performance across tasks such as code generation, multi-turn conversations, and mathematical reasoning \cite{naveed2023comprehensive}.
From ChatGPT \cite{openai2022gpt3-5,brown2020language} leading the global wave of AI applications to DeepSeek-R1 \cite{guo2025deepseek} redefining the landscape of reasoning LLMs, LLMs have emerged as core engines for powering next-generation intelligent interaction, content creation, and decision-making systems.

As dependence on LLM services intensifies, efficient orchestration and intelligent task allocation across multi-tier computing infrastructures become increasingly critical deployment challenges. Existing LLM serving paradigms, however, present significant limitations. On-device deployments offer enhanced privacy protection and eliminate transmission overhead, but are constrained by the limited local computational capabilities, supporting only lightweight LLMs \cite{liu2024mobilellm,hu2024minicpm} that compromise service quality for demanding tasks such as long-text generation and sophisticated reasoning. 
Centralized cloud-based deployment strategies \cite{fu2024serverlessllm,zhong2024distserve} deliver high-quality inference services but necessitate the transmission of full request prompts and contextual information across WANs to remote servers, potentially creating severe resource bottlenecks and network congestion during peak periods due to frequent service calls \cite{yang2024perllm,lin2023pushing}. 
While the model cascading-based multi-tier inference approach \cite{lebovitz2023efficient} partially addresses this challenge by distributing inference tasks across device, edge, and cloud nodes based on predefined hyperparameter thresholds. However, these static thresholds require fine-grained manual calibration for different task types, model configurations, and network conditions, severely limiting adaptability to dynamic real-world scenarios.

\color{black}
Given the heterogeneity and hierarchical nature inherent in multi-tier networks, we aim to answer the following research question: How can we design an LLM serving framework that intelligently leverages the computational resources across multiple tiers while maintaining service quality with reduced communication burden? This overarching question encompasses several critical sub-problems that must be addressed systematically. First, how should LLMs be distributed across device, edge, and cloud tiers to match node computational capabilities? Second, how can we effectively assess local inference quality to inform intelligent offloading decisions? Third, how can we develop an adaptive mechanism that leverages historical inference patterns to optimize offloading decisions without careful manual intervention? Addressing these interconnected issues requires a holistic design that balances resource utilization, service quality, and communication efficiency across the multi-tier infrastructure.
\color{black}

\textcolor{black}{In this paper, we propose RecServe, a recursive offloading framework for multi-tier networks designed to dynamically allocate LLM inference tasks based on customized confidence feedback. RecServe operates in a hierarchical LLM deployment scenario where progressively capable LLMs are positioned across device, edge, and cloud tiers. At the core of RecServe are two key components: task-specific confidence evaluation tailored to different NLP task categories, and a dynamic offloading strategy that leverages quantile interpolation over historical confidence scores to adaptively set decision thresholds.} When processing user requests, RecServe first performs local inference and computes a task-specific confidence score. Once this score surpasses the dynamically determined threshold, the local result is immediately returned. Otherwise, the task is recursively offloaded to the next higher tier until either a sufficiently confident result is obtained or the task reaches the cloud tier. This design ensures simpler tasks are handled locally on devices or at the edge, reserving cloud resources for computationally demanding requests. Our theoretical analysis establishes parameter bounds for offloading parameters under which RecServe achieves reduced expected communication burden and computational cost compared to centralized cloud-based deployment.
Experiments across eight datasets confirm that RecServe significantly outperforms the existing multi-tier LLM serving method, while reducing communication burden by more than 50\% compared to cloud-centric deployment.

In summary, the contributions of this paper are as follows:
\begin{itemize}
    \item We propose RecServe, a recursive offloading framework for LLM serving in multi-tier networks, which progressively offloads complex inference tasks to higher-capability tiers.
    \item We design a dynamic offloading strategy based on adaptive confidence thresholds, leveraging differentiated confidence evaluations tailored to distinct NLP tasks and quantile interpolation over historical confidence queues.
    \item We provide theoretical analysis for RecServe, establishing hyperparameter bounds for reducing expected communication burden and computational costs, respectively.
    \item We conduct extensive experiments on eight datasets. Results show that RecServe outperforms CasServe in both service quality and communication burden, and achieves over 50\% reduction in communication burden compared to cloud-based deployment.
\end{itemize}

\begin{table}[!t] 
\centering 
\caption{Main notations with descriptions.}
\label{main-notations}
\begin{adjustbox}{width=\linewidth}
\begin{tabular}{c|c}
\hline
\textbf{Notation} & \textbf{Description} \\
\hline
$n$ & \begin{tabular}[c]{@{}c@{}}Total number of tiers in hierarchical\\ deployment\end{tabular} \\
$M_i$ & LLM deployed on tier $i$ \\
$\tau$ & LLM task type \\
$x$ & Input text sequence \\
$y$ & Prediction generated by LLM \\


$H_{M,\tau}$ &  \begin{tabular}[c]{@{}c@{}}Historical confidence queue for LLM $M$ \\and task type $\tau$\end{tabular} \\
$k$ & \begin{tabular}[c]{@{}c@{}}Maximum capacity of the historical \\ confidence queue\end{tabular} \\
$C_{M,\tau}$ & \begin{tabular}[c]{@{}c@{}}Confidence score from LLM $M$ with task\\ type $\tau$\end{tabular} \\
$P_{M}(y=i|x)$ & \begin{tabular}[c]{@{}c@{}}Softmax probability assigned to class $i$\\ given input $x$\end{tabular} \\
$z_i$ & Logit corresponding to class $i$ \\
$\text{PPL}_{M}(y|x)$ & \begin{tabular}[c]{@{}c@{}}Perplexity of the output sequence $y$ \\given input $x$ by LLM $M$\end{tabular} \\
$\beta$ & \begin{tabular}[c]{@{}c@{}}Offloading parameter used for \\dynamic threshold computation\end{tabular} \\
$T_{M,\tau}$ & \begin{tabular}[c]{@{}c@{}}Dynamic offloading threshold for LLM $M$ \\and task type $\tau$\end{tabular} \\
$\mathcal{D}$ & Recursive offloading policy\\
$M^*$ & Final LLM that completes inference via $\mathcal{D}$\\
$p_i$ & \begin{tabular}[c]{@{}c@{}}Expected probability that a task is offloaded\\ from LLM $M_i$ to $M_{i+1}$\end{tabular} \\
$F_{M_{i},\tau}$ &\begin{tabular}[c]{@{}c@{}} Cumulative Distribution Function (CDF) of\\ confidence scores for $M_i$ and task type $\tau$\end{tabular} \\
$A$ & Availability check function\\
$\mathcal{D}_{ut}$ & \begin{tabular}[c]{@{}c@{}}Recursive offloading policy with \\ unavailability tolerance\end{tabular}\\

$M^{**}$ & \begin{tabular}[c]{@{}c@{}}Final LLM that completes inference\\ considering node unavailability via $\mathcal{D}_{ut}$\end{tabular}\\
\hline
\end{tabular}
\end{adjustbox}
\end{table}

\section{Preliminaries}
This section provides foundational knowledge necessary to understand the proposed framework. We will elaborate on the basics of LLM, the architecture of multi-tier networks, and the concept of task offloading. For clarity, a list of the main notations and descriptions used throughout this paper is provided in Table \ref{main-notations}.

\subsection{Large Language Models}
In recent years, LLMs have demonstrated remarkable capabilities in natural language understanding and generation, primarily driven by the Transformer module \cite{radford2019language} and the self-attention \cite{vaswani2017attention} mechanism. 
Mainstream approaches \cite{guo2025deepseek,yang2024qwen2} typically utilize a stack of attention layers augmented with positional encodings to form the backbone of LLMs, with the core computation represented as:
\begin{equation}
\text{Attention}(\mathbf{Q}, \mathbf{K}, \mathbf{V}) = \text{Softmax}\left(\frac{\mathbf{QK}^\top}{\sqrt{d_k}}\right) \cdot \mathbf{V},
\end{equation}
where $\mathbf{Q}$, $\mathbf{K}$, and $\mathbf{V}$ denote the query, key, and value matrices, respectively. $d_k$ is a scaling factor used to stabilize gradients during the training process.
To facilitate coherent and context-aware understanding and generation during inference, LLMs typically adopt pretraining objectives such as Masked Language Modeling \cite{devlin2019bert,liu2019roberta} or Causal Language Modeling (CLM) \cite{radford2018improving,radford2019language}. 
Taking widely adopted CLM as an example, an LLM $M$ optimizes its model parameters $\theta$ by maximizing the probability of generating the current token conditioned on all previous tokens:
\begin{equation}
\max_{\theta} \ \mathrm{P}_{M}(x_1, \ldots, x_m) = \max_{\theta} \ \prod_{i=1}^{m} \mathrm{P}_{M}(x_i \mid x_1, \ldots, x_{i-1}),
\end{equation}
where $x = (x_1, x_2, \dots, x_m)$ is the input text sequence, $x_i$ represents the $i$-th token in $x$, and $m$ denotes the sequence length. Besides, $\mathrm{P}(x_i \mid x_1, \dots, x_{i-1})$ represents the probability of predicting the $i$-th token given the preceding tokens. Accordingly, the training objective of CLM is to minimize the negative log-likelihood loss $\mathcal{L}_M$ as follows:
\begin{equation}
\min_{\theta} \ \mathcal{L}_M = \min_{\theta} -\sum_{i=1}^{n} \log \mathrm{P}_{M}(x_i \mid x_1, \ldots, x_{i-1}).
\end{equation}
Beyond pretraining, LLMs typically undergo additional processes such as task-specific fine-tuning \cite{hu2022lora,houlsby2019parameter} or reinforcement learning from human feedback \cite{ouyang2022training}, bridging the gap between general language modeling capabilities and practical application requirements.
During inference, LLMs exhibit the ability to handle multiple task formats \cite{raffel2020exploring}, which can be broadly categorized into Seq2Class (Sequence-to-Class) and Seq2Seq (Sequence-to-Sequence) categories. The former refers to tasks that map an input sequence to a single class label, and the latter involves generating an output sequence conditioned on the input sequence. These capabilities empower LLMs to serve as foundational models across a wide range of downstream tasks in real-world applications.

\subsection{Multi-tier Networks}
Multi-tier network integrates heterogeneous computing resources through a vertically tiered design paradigm, which typically organizes computing nodes into the device tier, the edge node tier, and the cloud data center tier \cite{duan2023,yang2023hierarchical,wang2024end,wu2025beyond}. This hierarchical division is motivated by significant differences across tiers in computational capacity and communication latency \cite{shi2016edge,wu2024agglomerative}.
At the device tier, embedded chips with limited computational power are deployed, primarily suited for sensor data collection and lightweight inference tasks. The edge node tier incorporates more powerful computing infrastructures that support real-time service coordination and handle moderately complex computational workloads. In contrast, the cloud data center tier is equipped with powerful GPU or TPU clusters but with relatively high communication latencies.

In the context of LLM serving, multi-tier networks can potentially support distributed inference and enable dynamic communication-precise trade-offs via intelligent tiered deployment strategies. By integrating flexible task-routing pathways within this architecture, on-demand LLM inference workload distribution becomes feasible. As a result, most lightweight requests can be effectively processed at the device or edge tiers, while only a small subset of computationally intensive tasks necessitate cloud-based computation. Such hierarchical cooperation substantially prevents congestion at the cloud tier, reduces data transmission across public networks, and minimizes communication overheads with maintenance on service quality.

\subsection{Task Offloading}
Task offloading refers to the process of migrating computational tasks from local nodes to remote ones to optimize system performance and resource utilization \cite{islam2021survey,saeik2021task}. In a typical two-tier computing system, the objective of task offloading $\mathcal{C}_{system}$ can be formulated as an optimization problem minimizing total execution cost:
\begin{equation}
\min_{\tau \in \{0, 1\}} \ \mathcal{C}_{system} =
(1 - \tau) \cdot \mathcal{C}_{local} +
\tau \cdot (\mathcal{C}_{comm} + \mathcal{C}_{remote}),
\end{equation}
where $\tau$ is the offloading decision variable, with $\tau=0$ denoting local execution and $\tau=1$ representing offloading to a remote node. $\mathcal{C}_{{local}}$ denotes the cost of local execution, $\mathcal{C}_{{comm}}$ represents the communication overhead incurred during offloading, and $\mathcal{C}_{{remote}}$ corresponds to the cost of remote execution. These costs are primarily determined by task characteristics, computational capacity, and network conditions \cite{dong2024task}.
Such a formulation encapsulates the core principle of task offloading: dynamically balancing between local execution and remote execution plus communication, and selecting the path with the lower overall cost.

In practical LLM serving scenarios, $\mathcal{C}_{{local}}$ corresponds to the accuracy loss incurred by executing a lightweight LLM locally. $\mathcal{C}_{{remote}}$ reflects the computational resource consumption required to process the task using a large LLM on a remote node, and is closely related to inference quality. $\mathcal{C}_{{comm}}$ is heavily influenced by transmitted sequence length. Based on this system model, task-aware or network-aware strategies can be employed to dynamically determine the offloading path, facilitating collaborative task execution and resource adaptation in multi-tier networks.

\begin{figure*}[!t]
    \centering
    \begin{minipage}{\linewidth}
        \centering
        \includegraphics[width=\textwidth]{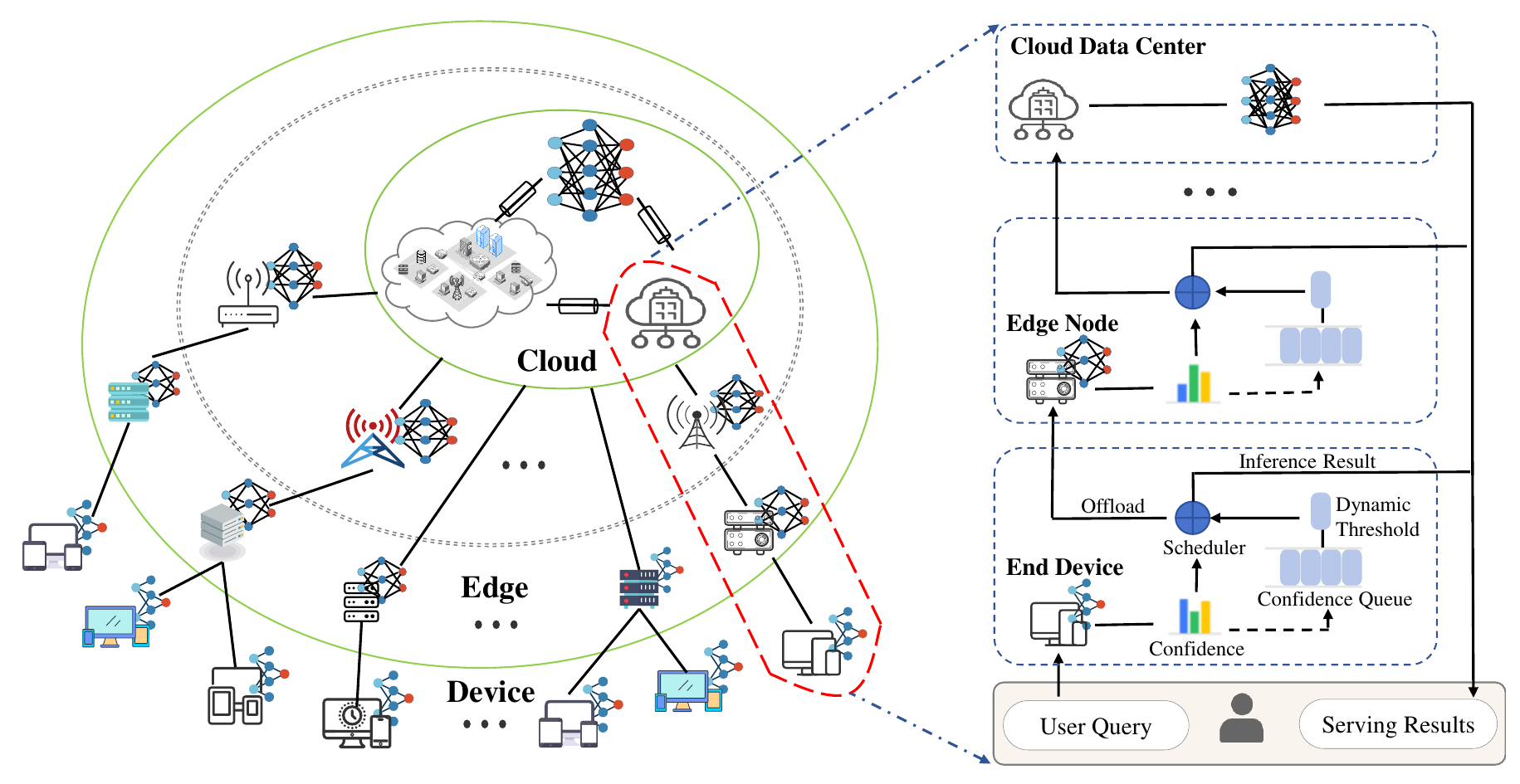}
        \caption{Schematic diagram of RecServe.}
        \label{framework}
    \end{minipage}\par\medskip
\end{figure*}

\section{Recursive Offloading}
This section details the proposed RecServe framework. We will begin with an overview of the RecServe framework, followed by explanations of its core components: the historical confidence queue, task-specific confidence evaluation mechanisms, and the recursive offloading strategy with dynamic decision thresholds. Finally, we provide the formal description of RecServe.

\subsection{Framework Overview}
The schematic diagram of RecServe is illustrated in Fig. \ref{framework}, which spans across end devices, edge nodes, and cloud data centers. Specifically, RecServe deploys LLMs with progressively increasing model sizes and inference capabilities at the end, edge, and cloud, respectively, to align with the heterogeneous computational resources of different tiers.
Each computational node maintains a historical confidence queue to store the computed confidence scores previously processed by its local LLM. To support a broad range of Natural Language Processing (NLP) tasks, RecServe incorporates task-specific confidence evaluation mechanisms: peak softmax probability for Sequence to Class (Seq2Class) tasks and normalized perplexity for Sequence to Sequence (Seq2Seq) tasks.
Based on a dynamic offloading strategy relying on a sliding window of historical confidence scores, RecServe continuously tracks the distribution of past confidence levels and dynamically adjusts its offloading threshold. 
This adaptive mechanism enables each node to decide whether to process a task locally or escalate it to a higher-tier node based on the quality of the current inference. As only a small subset of complex tasks are transmitted over the network, the majority of tasks are processed locally without cross-tier communication. This design enables RecServe to distribute the inference load across different tiers, reducing pressure on cloud data centers and mitigating resource congestion during peak periods, while also achieving an effective balance between communication burden and service performance.

In the following subsections, we will provide a detailed explanation of the constituent parts of RecServe, including the historical confidence queue, the task-specific confidence evaluation mechanism, and the recursive offloading strategy. A formal description of the execution procedure of RecServe is presented at the end of this section.

\subsection{Historical Confidence Queue}
By maintaining an up-to-date confidence history, each computational node can effectively analyze historical inference quality, estimate the complexity distribution of incoming tasks, and dynamically adjust offloading thresholds to make informed decisions about whether to handle tasks locally or forward them to higher-tier nodes. To facilitate this, we design a sliding-window-based historical confidence queue to capture the confidence information of LLMs in the past inference process, which can be formally defined as:
\begin{equation}
H_{M, \tau} = \{C_{M, \tau}(x_1), C_{M, \tau}(x_2), \ldots, C_{M, \tau}(x_k)\},
\label{queue-define}
\end{equation}
where $H_{M, \tau}$ denotes the historical confidence queue for LLM $M$ and task type $\tau$, $C_{M, \tau}(x_i)$ denotes the confidence score for input $x_i$, and $k$ represents the maximum queue capacity. Each node initializes its historical confidence queue as empty and maintains it following the First-In-First-Out (FIFO) policy, so as to ensure the queue reflects the most recent $k$ predictions:
\begin{equation}
\begin{array}{l}
H_{M, \tau}^{\text{new}} =\\
\begin{cases}
H_{M, \tau} \cup \{C_{M, \tau}(x_{\text{new}})\}, & \text{if } |H_{M, \tau}| < k \\
(H_{M, \tau} - \{C_{M, \tau}(x_1)\}) \cup \{C_{M, \tau}(x_{\text{new}})\}, & \text{if } |H_{M, \tau}| = k,
\end{cases}
\end{array}
\label{queue-update}
\end{equation}
where $x_{\text{new}}$ denotes the latest input and $H_{M, \tau}^{\text{new}}$ is the updated queue.

\subsection{Task-Specific Confidence Evaluation}
The confidence measure associated with an LLM's output should be adapted to the characteristics of the current inference task, as it plays a critical role in determining the reliability and suitability of offloading decisions.
Accordingly, we employ task-specific confidence evaluation mechanisms for Seq2Class and Seq2Seq tasks in LLM serving.

For Seq2Class tasks, we define output confidence as the maximum softmax probability. First, given an input $x$, the softmax probability assigned to each class $i$ is computed as follows:
\begin{equation}
P_M(y=i|x) = \frac{\exp(z_i)}{\sum_{j=1}^{C} \exp(z_j)},
\label{seq2class1}
\end{equation}
where $z_i$ is the logit corresponding to class $i$ output by LLM $M$, and $C$ is the total number of classes. Then, the confidence score can be formulated as follows:
\begin{equation}
C_{M,\text{Seq2Class}}(x) = \max_{i \in [1,C]} P_M(y=i|x).
\label{seq2class2}
\end{equation}
In Seq2Class scenarios, the primary goal is to identify the most probable category for a given input text. Therefore, the maximum softmax probability naturally provides a direct measure of the LLM's certainty in its prediction \cite{pearce2021understanding}. A higher confidence score indicates greater LLM certainty, thereby reducing the necessity of offloading the task to alternative resources.

For Seq2Seq tasks, we adopt perplexity \cite{radford2019language} with customized normalization as the confidence metric, aiming to assess the coherence and consistency of the entire token sequence.
Consider the output text $y$ as a generated sequence comprising individual tokens:
\begin{equation}
y = (t_1, t_2, \ldots, t_L)
\label{seq2seq1}
\end{equation}
where $t_i$ denotes the $i$-th token generated by the LLM, and $L$ is the sequence length. The perplexity is computed  based on the average negative log-likelihood of the token sequence:
\begin{equation}
\text{PPL}_M(y|x) = \exp\left( -\frac{1}{L} \sum_{i=1}^{L} \log P_M(t_i|t_1, \ldots, t_{i-1}, x) \right),
\label{seq2seq2}
\end{equation}
where $P_M(t_i|t_1, \ldots, t_{i-1}, x)$ denotes the softmax probability assigned by LLM $M$ to token $t_i$ given the input $x$ and the previous generated tokens. Specifically,
\begin{equation}
P_M(t_i|t_1, \ldots, t_{i-1}, x) = \frac{\exp(z_{t_i})}{\sum_{j=1}^{|V|} \exp(z_{V_j})},
\label{seq2seq3}
\end{equation}
where $z_{t_i}$ is the logit associated with token $t_i$, and $V$ is the vocabulary set.
Since lower perplexity values reflect higher confidence and greater coherence in sequence generation, we normalize perplexity into a confidence score in the range $(0,1)$ as follows:
\begin{equation}
C_{M,\text{Seq2Seq}}(x) = \frac{1}{1 + \text{PPL}_M(y|x)}.
\label{seq2seq4}
\end{equation}
This normalization step provides an intuitive interpretation of confidence scores, where larger values indicate stronger LLM certainty.

\subsection{Recursive Offloading with Dynamic Decision Thresholds}
Building upon the historical confidence queue and task-adaptive confidence evaluation, RecServe implements a strategy with dynamic decision thresholds. Unlike the prior method that depends on fixed thresholds and tier-specific hyperparameters \cite{lebovitz2023efficient}, our approach dynamically adjusts the offloading threshold based on statistical characteristics derived from the historical confidence queue, effectively reducing unnecessary communication while preserving inference quality.
Formally, given the historical confidence queue 
$H_{M, \tau}$, we first obtain a sorted version of this queue in ascending order:
\begin{equation}
H_{M,\tau}^{\text{sorted}} = \{c_{(1)}, c_{(2)}, \ldots, c_{(k)}\}
\label{threthold-1}
\end{equation}
where $c_{(i)}$ denotes the $i$-th confidence score in the sorted list, satisfying:
\begin{equation}
c_{(1)} \leq c_{(2)} \leq \ldots \leq c_{(k)}
\label{threthold-2}
\end{equation}
Based on the sorted queue, RecServe defines the dynamic threshold $T_{M, \tau}(\beta)$ as the $\beta$-th quantile of the confidence distribution in $H_{M, \tau}$, with $\beta \in (0,1)$. Specifically, the threshold computation employs linear interpolation as follows:
\begin{equation}
T_{M,\tau}(\beta) = c_{(\lfloor r \rfloor + 1)} \cdot (1 - (r - \lfloor r \rfloor)) + c_{(\lceil r \rceil + 1)} \cdot (r - \lfloor r \rfloor)
\label{threthold-3}
\end{equation}
where $r = \beta \cdot (k - 1)$. A smaller $\beta$ results in a stricter threshold, reducing the proportion of offloaded tasks, while a larger $\beta$ relaxes the threshold, leading to more frequent offloading.
This dynamically computed threshold guides the recursive decision-making process for task offloading at each computational node. For a given non-top-tier LLM $M$, the input text $x$ is processed locally if the confidence score satisfies:
\begin{equation}
    C_{M, \tau}(x) \geq T_{M, \tau}(\beta).
    \label{local-process}
\end{equation}
Otherwise, the task is recursively offloaded to the next higher-tier LLM $M'$. As a result, the recursive offloading decision policy  $\mathcal{D}(x, M, \tau)$ can be formalized as:
\begin{equation}
\begin{array}{l}
\mathcal{D}(x, M, \tau) =\\
\begin{cases}
M, & \text{if } C_{M, \tau}(x) \geq T_{M, \tau}(\beta) \\
\mathcal{D}(x, M', \tau), & \text{if } M \neq M_n \text{ and } C_{M, \tau}(x) < T_{M, \tau}(\beta),
\end{cases}
\end{array}
\label{recur-offload}
\end{equation}
where $M'$ represents the LLM on the upstream computational node of $M$. The recursion terminates either when an LLM achieves sufficient confidence or when the task reaches the highest-tier computational node. Consequently, the final inference output produced by RecServe is given by:
\begin{equation}
y = M^*(x), M^*=\mathcal{D}(x, M_1, \tau),
\end{equation}
where $M^*(x)$ is the final LLM that completes inference. $M_1$ denotes the initial processing node, typically an end device that close to users.




\begin{algorithm}[t]
\caption{Recursive Offloading for LLM Serving}
\SetKwFunction{FMain}{RecServe}
\SetKwFunction{FOffload}{TaskOffloading}

\SetKwProg{Fn}{Function}{:}{}
\Fn{\FMain{$x, \tau, \beta, \{M_1, M_2, \dots, M_n\}$}}{
    $y \leftarrow$ \FOffload{$x, M_1, \tau, \beta$}\\
    \Return $y$
}

\Fn{\FOffload{$x, M_i, \tau, \beta$}}{
    $(\text{logits}, y) \leftarrow M_i(x)$

    \uIf{$\tau = \text{Seq2Class}$}{
        Compute $C_{M_i, \tau}(x)$ following Eqs. (\ref{seq2class1},\ref{seq2class2})
    }
    \Else{
        Compute $C_{M_i, \tau}(x)$ following Eqs. (\ref{seq2seq1},\ref{seq2seq2},\ref{seq2seq3},\ref{seq2seq4})
    }

    Update $H_{M, \tau}$ following Eqs. (\ref{queue-define},\ref{queue-update})\\
    Compute $T_{M,\tau}(\beta)$ following Eqs. (\ref{threthold-1},\ref{threthold-2},\ref{threthold-3})\\
    \uIf{$C_{M_i, \tau}(x) \geq T_{M,\tau}(\beta)$ \textbf{or} $M_i$ is $M_n$}{
        \uIf{$M_i$ is not $M_1$}{
        Transmit $y$ from $M_i$ to $M_{i-1}$
        }
        \Return $y$
    }
    \Else{
        Transmit $x$ from $M_i$ to $M_{i+1}$\\
        $y \leftarrow$ \FOffload{$x, M_{i+1}, \tau, \beta$}
        \Return $y$
    }
}
\label{recserve-algorithm}
\end{algorithm}

\subsection{Formal Description of RecServe}
As illustrated in Algorithm \ref{recserve-algorithm} and Fig. \ref{framework}, RecServe is deployed within a multi-tiered network architecture comprising end devices, edge nodes, and cloud data centers. The proposed framework employs a recursive offloading strategy based on a hierarchical evaluation of confidence scores guided by dynamic thresholds.


The serving process of RecServe initiates at end devices equipped with a lightweight local LLM $M_1$. For each input text $x$ with task type $\tau$, the recursive offloading function $\text{TaskOffloading}(x, M_i, \tau, \beta)$ (as detailed in Algorithm \ref{recserve-algorithm}) is invoked. At each computational tier, the LLM $M_i$ first performs local inference to generate its prediction output $y$ and computes the corresponding confidence score $C_{M_i,\tau}(x)$ using either maximum softmax probability for Seq2Class tasks (Eqs. (\ref{seq2class1},\ref{seq2class2})) or normalized perplexity for Seq2Seq tasks (Eqs. (\ref{seq2seq1},\ref{seq2seq2},\ref{seq2seq3},\ref{seq2seq4})).
Subsequently, $M_i$ updates its historical confidence queue $H_{M_i,\tau}$ (Eqs. (\ref{queue-define}, \ref{queue-update})) and computes a dynamic offloading threshold $T_{M_i,\tau}(\beta)$ based on the interpolated $\beta$-quantile of the updated confidence history (Eqs. (\ref{threthold-1},\ref{threthold-2},\ref{threthold-3})). The inference result $y$ is finalized at the current tier if either the computed confidence $C_{M_i,\tau}(x)$ meets or exceeds the threshold (Eq. (\ref{local-process})) or if the current LLM $M_i$ represents the top-tier cloud LLM $M_n$. In such cases, the result is directly returned or propagated back to lower-tier nodes. Otherwise, the input $x$ is recursively offloaded to the next-tier LLM $M_{i+1}$ for further inference (Eq. (\ref{recur-offload})). This recursive procedure continues until a sufficiently confident result is obtained or the highest-tier cloud LLM is reached.

By leveraging historical feedback and adaptive confidence thresholds, RecServe ensures that only challenging and uncertain tasks are transmitted upward through the network, while the majority of requests are efficiently processed at local computational nodes.
Through this design, RecServe reduces the load on cloud resources to mitigate potential congestion while achieving a fine-grained balance between inference quality and communication overhead.




\section{Theoretical Analysis for RecServe}
This section provides the theoretical support for RecServe. We will present an approximation of the communication burden and an analysis of the computational costs associated with the proposed RecServe under specific assumptions.
\subsection{Communication Burden Approximation}
\label{comm-approx}
To estimate the total communication burden incurred by RecServe across multi-tier networks, we analyze both the probability of task offloading between hierarchical tiers and the transmission cost associated with each offloading event.

Assume RecServe deploy $n$ LLMs of increasing scale across end, edge, and cloud tiers, denoted as $M_1, M_2, \ldots, M_n$, where $M_1$ is the smallest LLM and $M_n$ is the most powerful one. Each task is initially processed by $M_1$ and may be offloaded to higher-tier LLMs based on its confidence score. Let $x$ denote an input text of length $|x|$ and $y$ denote the ultimate inference result of length $|y|$. Communication burden occurs primarily when transmitting $x$ to a higher-tier node and returning $y$ to the lower-tier node.
Each time a task is offloaded from $M_i$ to $M_{i+1}$, the input $x$ is uploaded, incurring a communication burden of $|x|$ for both the sending and receiving nodes. Upon task completion at a certain tier $M^*$, the inference result $y$ is propagated back through each intermediate tier from $M^*$ to $M_1$, incurring a communication burden of $|y|$ at each node along the return path.

Let $p_i$ denote the expected probability that a task is offloaded from $M_i$ to $M_{i+1}$. According to the recursive offloading strategy in RecServe, $p_i$ is defined as the probability that the task's confidence score at $M_i$ falls below a quantile threshold $\beta$ computed from the historical confidence queue $H_{M_i, \tau}$, that is:
\begin{equation}
p_i = P(C_{M_i, \tau}(x) < T_{M_i, \tau}(\beta))
\end{equation}
We can approximate $p_i$ under the following assumptions.

\noindent
\textbf{Assumption 1: Independent and Identically Distributed Confidence Scores.} The confidence score $C_{M_i, \tau}(x)$ for the current task is assumed to be independently and identically distributed with respect to the historical confidence scores stored in the queue. Let $F_{M_i, \tau}(c)$ denote the Cumulative Distribution Function (CDF) of $C_{M_i, \tau}(x)$:
\begin{equation}
F_{M_i, \tau}(c) = P(C_{M_i, \tau}(x) \leq c).
\end{equation}
We assume that $F_{M_i, \tau}(c)$ is determined solely by the task difficulty and the capacity of the current LLM $M_i$.

\noindent
\textbf{Assumption 2: Sufficient Sample in Historical Queue.} The historical confidence queue contains a sufficiently large number of samples such that the threshold $T_{M_i, \tau}(\beta)$ closely approximates the $\beta$-quantile of $F_{M_i, \tau}(c)$:
\begin{equation}
P(C_{M_i, \tau}(x) \leq T_{M_i, \tau}(\beta)) \approx \beta.
\end{equation}

\noindent
\textbf{Assumption 3: Monotonic Continuity of CDF.}  
The CDF $F_{M_i, \tau}(c)$ is monotonically continuous everywhere.

\noindent
\textbf{Assumption 4: Independence Between Confidence and Input Length.}  
The text length $|x|$ is assumed to be uncorrelated with the confidence score $C_{M, \tau}(x)$ produced during LLM inference.

\noindent
\textbf{Assumption 5: Output Length Distribution Invariance Across Tiers}: The distribution of output lengths $|y|$ remains consistent across different processing tiers.

According to Assumption 1, the confidence scores in the historical queue $H_{M_i, \tau}$ are sufficient independent samples drawn from the distribution $F_{M_i, \tau}(c)$. Due to the monotonicity nature of $F_{M_i, \tau}(c)$ following Assumption 3, we can define the inverse function $F_{M_i, \tau}^{-1}(\beta)$, satisfying:
\begin{equation}
F_{M_i, \tau}(F_{M_i, \tau}^{-1}(\beta)) = \beta.    
\end{equation}
This implies:
\begin{equation}
    P(C_{M_i, \tau}(x) \leq F_{M_i, \tau}^{-1}(\beta)) = \beta,
\end{equation}
indicating that $F_{M_i, \tau}^{-1}(\beta)$ represents the theoretical $\beta$-quantile of the distribution $F_{M_i, \tau}(c)$. By Assumption 2, we approximate:
\begin{equation}
P(C_{M_i, \tau}(x) \leq F_{M_i, \tau}^{-1}(\beta)) = \beta \approx P(C_{M_i, \tau}(x) \leq T_{M_i, \tau}(\beta)).
\end{equation}
Hence, we obtain the approximation:
\begin{equation}
    T_{M_i, \tau}(\beta) \approx F_{M_i, \tau}^{-1}(\beta),
\end{equation}
which further implies:
\begin{equation}
    p_i = P(C_{M_i, \tau}(x) < T_{M_i, \tau}(\beta)) \approx P(C_{M_i, \tau}(x) < F_{M_i, \tau}^{-1}(\beta)).
\end{equation}
Under Assumption 3 where $F_{M_i, \tau}(c)$ is continuous at $F_{M_i, \tau}^{-1}(\beta)$, we have:
\begin{equation}
    \lim_{c \uparrow F_{M_i, \tau}^{-1}(\beta)} F_{M_i, \tau}(c) = F_{M_i, \tau}(F_{M_i, \tau}^{-1}(\beta)) = \beta.
\end{equation}
Therefore, the jump at point $F_{M_i, \tau}^{-1}(\beta)$ is zero, which means:
\begin{equation}
    P(C_{M_i, \tau}(x) = F_{M_i, \tau}^{-1}(\beta)) = 0,
\end{equation}
which leads to:
\begin{equation}
    P(C_{M_i, \tau}(x) < F_{M_i, \tau}^{-1}(\beta)) = P(C_{M_i, \tau}(x) \leq F_{M_i, \tau}^{-1}(\beta)) = \beta.
\end{equation}
Therefore, we can conclude that:
\begin{equation}
    p_i \approx \beta.
    \label{approx-pi}
\end{equation}
Based on the approximation in Eq. (\ref{approx-pi}), we derive the probability of task completion at each tier and the corresponding communication burden. According to the serving process of RecServe, tasks are initially processed at $M_1$ and are offloaded to higher tiers only when the confidence is insufficient. Let $P^C(M)$ denote the probability that a task is completed at LLM $M$. Then, the probability of completing the inference task at LLM $M_i$ at each tier can be approximated as follows.
\begin{itemize}
    \item Task completed at at $M_1$ without offloading:
    \begin{equation}
        P^C(M_1) = 1 - p_1 \approx 1 - \beta.
    \end{equation}
    \item Task completed at $M_i$, $i \in \{2, \ldots, n-1\}$ after $i-1$ offloads without further offloading to higher tiers:
    \begin{equation}
        P^C(M_i) = \prod_{j=1}^{i-1} p_j \cdot (1 - p_i) \approx \beta^{i-1} (1 - \beta).
    \end{equation}
    \item Task completed at $M_n$, after being offloaded to the highest tier:
    \begin{equation}
        P^C(M_n) = \prod_{i=1}^{n-1} p_i \approx \beta^{n-1}.
    \end{equation}
\end{itemize}
Following Assumptions 4 and 5, the expected communication burden for each type of inference can be derived as follows:
\begin{itemize}
    \item When the task is completed at $M_1$, no communication burden is incurred.
    \item When the task is completed at $M_i$, $i \in \{2, \ldots, n-1\}$, the input $x$ is uploaded $i-1$ times, and the output $y$ is downloaded $i-1$ times. Therefore, the total communication burden introduced to the pathway computing nodes is:
    \begin{equation}
        2(i-1)(|x| + |y|).
    \end{equation}
    \item When the task is completed at $M_n$ in the highest tier, the input $x$ is uploaded $n-1$ times, and the output $y$ is downloaded $n-1$ times. Thus, the incurred total communication burden is computed as:
    \begin{equation}
        2(n-1)(|x| + |y|).
    \end{equation}
\end{itemize}
Thus, the expected communication burden under RecServe $E[\text{Comm-RecServe}]$ is as follows:
\begin{equation}
    \begin{array}{l}
\;\;\;\;E[\text{Comm-RecServe}] \\=\sum_{i=2}^{n-1}P^C(M_i)\cdot2(i-1)(|x| + |y|)\\ \;\;\;\; +P^C(M_n)\cdot 2(n-1)(|x| + |y|)\\ \approx 2\sum_{i=2}^{n-1} (i-1)(|x| + |y|) \times\beta^{i-1} (1 - \beta) \\ \;\;\;\; + 2(n-1)(|x| + |y|) \times \beta^{n-1}\\\approx2(|x| + |y|) \left[ (1 - \beta) \sum_{i=2}^{n-1} (i-1) \beta^{i-1} + (n-1) \beta^{n-1} \right]\\\approx2(|x| + |y|) \left[ \beta \frac{1 - (n-1) \beta^{n-2} + (n-2) \beta^{n-1}}{1 - \beta} \right. \\ \;\;\;\; \left.+ (n-1) \beta^{n-1} \right]
\end{array}
\end{equation}
In particular, for a three-tier device-edge-cloud network architecture, i.e., $n = 3$, we have:

\begin{equation}
\begin{array}{l}
\;\;\;\;E[\text{Comm-RecServe}] \\ \approx 2(|x| + |y|) \left[ \frac{\beta(1 - 2\beta + \beta^2)}{1 - \beta} + 2\beta^2 \right] \\
\approx 2(|x| + |y|) \cdot \beta(1 + \beta).
\end{array}
\end{equation}
For comparison, consider the cloud-centric LLM serving scheme (CloudServe), where all task requests are sent directly to the cloud data center. The expected total communication burden for this serving paradigm is:
\begin{equation}
    E[\text{Comm-CloudServe}] = 2(|x| + |y|).
\end{equation}
Therefore, the ratio of communication burden between the two serving paradigms can be computed as:
\begin{equation}
    \frac{E[\text{Comm-RecServe}]}{E[\text{Comm-CloudServe}]} = \beta(1 + \beta).
\end{equation}
For RecServe to be more communication-efficient than cloud-only serving, we require:
\begin{equation}
    \frac{E[\text{Comm-RecServe}]}{E[\text{Comm-CloudServe}]} \in (0, 1) \wedge \beta \in (0, 1).
\end{equation}
This yields:
\begin{equation}
\beta \in \left(0, \frac{\sqrt{5}-1}{2} \right)
\end{equation}
Within this range, RecServe is expected to be more communication-efficient than the cloud-only serving paradigm under our assumptions.

\subsection{Computation Cost Analysis}
For simplicity, our analysis focuses solely on the computational cost of LLM inference, excluding other factors such as data communication overhead between neighboring nodes and the management of historical confidence queues.
Let the inference cost of the $n$ hierarchical LLMs $M_1, M_2, \ldots, M_n$ be denoted as $\text{Cost}_1, \text{Cost}_2, \ldots, \text{Cost}_n$, respectively, satisfying $\text{Cost}_1 < \text{Cost}_2 < \cdots < \text{Cost}_n$. Based on the assumptions in section \ref{comm-approx}, the task completion probabilities at $M_1$, $M_i$ $(2 \leq i \leq n-1)$, and $M_n$ can be approximated as $1 - \beta$, $\beta^{i-1}(1 - \beta)$, and $\beta^{n-1}$, respectively. Thus, the expected computation cost of RecServe $E[\text{Comp-RecServe}]$ is given by:
\begin{equation}
\begin{array}{l}
\;\;\;\;E[\text{Comp-RecServe}]\\
= P^C(M_1)\cdot\text{Cost}_1 + \sum_{i=2}^{n-1} P^C(M_i)\cdot\sum_{j=1}^{i} \text{Cost}_j  \\ \;\;\;\; + P^C(M_n)\cdot\sum_{j=1}^{n} \text{Cost}_j \\
\approx (1 - \beta)\cdot\text{Cost}_1 + \sum_{i=2}^{n-1} \beta^{i-1}(1 - \beta)\cdot\sum_{j=1}^{i} \text{Cost}_j \\ \; \;\;\;+ \beta^{n-1}\cdot\sum_{j=1}^{n} \text{Cost}_j
\end{array}
\end{equation}
In a three-tier device-edge-cloud architecture, we can specify $\text{Cost}_1 = \text{Cost}_\text{device}$, $\text{Cost}_2 = \text{Cost}_\text{edge}$, and $\text{Cost}_3 = \text{Cost}_\text{cloud}$, yielding:

\begin{equation}
\begin{array}{l}
\;\;\;\;E[\text{Comp-RecServe}]\\
\approx (1 - \beta)\cdot\text{Cost}_\text{device} + \beta(1 - \beta)\cdot(\text{Cost}_\text{device} + \text{Cost}_\text{edge}) \\ \;\;\;\;+ \beta^2\cdot(\text{Cost}_\text{device} + \text{Cost}_\text{edge} + \text{Cost}_\text{cloud}) \\
\approx \text{Cost}_\text{device} + \beta \cdot \text{Cost}_\text{edge} + \beta^2 \cdot \text{Cost}_\text{cloud}
\end{array}
\end{equation}
For comparison, the expected computation cost of cloud-only inference $E[\text{Comp-CloudServe}]$ is:
\begin{equation}
E[\text{Comp-CloudServe}] = \text{Cost}_\text{cloud}
\end{equation}
Therefore, the relative computation cost of RecServe compared to cloud-only inference can be expressed as:
\begin{equation}
\begin{array}{l}
 \;\;\;\; \frac{E[\text{Comp-RecServe}]}{E[\text{Comp-CloudServe}]}  \\ \approx \frac{\text{Cost}_\text{device} + \beta \cdot \text{Cost}_\text{edge} + \beta^2 \cdot \text{Cost}_\text{cloud}}{\text{Cost}_\text{cloud}}
\end{array}
\end{equation}
For RecServe to outperform cloud-only inference in terms of computational efficiency, we require:
\begin{equation}
\frac{E[\text{Comp-RecServe}]}{E[\text{Comp-CloudServe}]} < 1
\end{equation}
Therein, we can infer that:
\begin{equation}
    \adjustbox{width=0.48\textwidth}{\(\beta \in \left(0, \frac{-\text{Cost}_\text{edge} + \sqrt{\text{Cost}_\text{edge}^2 + 4\text{Cost}_\text{cloud}(\text{Cost}_\text{cloud} - \text{Cost}_\text{device})}}{2 \text{Cost}_\text{cloud}}\right)\)}
\end{equation}
Within this range, RecServe is expected to incur lower computation cost than cloud-only inference.

\begin{table*}[!t]
        \centering
        \caption{Inference accuracy vs communication burden in Seq2Class tasks.}
        \renewcommand{\arraystretch}{0.95}
        \label{seq2class-exp} 
        \begin{adjustbox}{width=0.88\textwidth}
        \begin{tabular}{c|l|c|cccc}
\hline
\multirow{11}{*}{\textbf{IMDB}}             & \multicolumn{1}{c|}{\multirow{2}{*}{\textbf{Method}}} & \multirow{2}{*}{\textbf{Accuracy (\%)}} & \multicolumn{4}{c}{\textbf{Communication Burden (B)}}            \\
                                            & \multicolumn{1}{c|}{}                                 &                                         & \textbf{Device} & \textbf{Edge} & \textbf{Cloud} & \textbf{Total} \\ \cline{2-7} 
                                            & EndServe                                              & 90.92                                   & -            & -             & -              & -              \\
                                            & EdgeServe                                             & 92.29                                   & 30.41M       & 30.41M        & -              & 60.82M         \\
                                            & CloudServe                                            & 94.25                                   & 30.41M       & -             & 30.41M         & 60.82M         \\
                                            & ColServe($\alpha=0.2$)                                    & 91.32                                   & 6.09M        & 7.33M         & 1.23M          & 14.65M         \\
                                            & CasServe(setting 1)                         & 92.33                                   & 3.92M        & 4.40M         & 0.48M          & 8.80M          \\
                                            & RecServe($\beta=0.1$)                                    & 92.35                                   & 3.73M        & 4.15M         & 0.42M          & 8.30M          \\
                                            & ColServe($\alpha=0.5$)                                    & 91.88                                   & 15.08M       & 22.72M        & 7.64M          & 45.44M         \\
                                            & CasServe(setting 2)                         & 93.32                                   & 12.27M       & 15.35M        & 3.08M          & 30.70M         \\
                                            & RecServe($\beta=0.3$)                                    & 93.74                                   & 10.97M       & 14.61M        & 3.64M          & 29.22M         \\ \hline
\multirow{11}{*}{\textbf{SST-2}}             & \multicolumn{1}{c|}{\multirow{2}{*}{\textbf{Method}}} & \multirow{2}{*}{\textbf{Accuracy (\%)}} & \multicolumn{4}{c}{\textbf{Communication Burden (B)}}          \\
                                            & \multicolumn{1}{c|}{}                                 &                                         & \textbf{Device} & \textbf{Edge} & \textbf{Cloud} & \textbf{Total} \\ \cline{2-7} 
                                            & EndServe                                              & 91.17                                   & -            & -             & -              & -              \\
                                            & EdgeServe                                             & 94.38                                   & 89.80K       & 89.80K        & -              & 179.60K        \\
                                            & CloudServe                                            & 95.76                                   & 89.80K       & -             & 89.80K         & 179.60K        \\
                                            & ColServe($\alpha=0.2$)                                    & 91.74                                   & 18.77K       & 22.60K        & 3.83K          & 45.20K         \\
                                            & CasServe(setting 3)                         & 93.35                                   & 12.38K       & 13.58K        & 1.20K          & 27.16K         \\
                                            & RecServe($\beta=0.1$)                                    & 93.58                                   & 11.67K       & 12.97K        & 1.30K          & 25.94K         \\
                                            & ColServe($\alpha=0.5$)                                    & 92.43                                   & 43.97K       & 63.18K        & 19.21K         & 126.36K        \\
                                            & CasServe(setting 4)                     & 94.95                                   & 37.29K       & 44.57K        & 7.28K          & 89.14K         \\
                                            & RecServe($\beta=0.3$)                                    & 95.30                                   & 29.76K       & 40.56K        & 10.80K         & 81.12K         \\ \hline
\multirow{11}{*}{\textbf{Rotten Tomatoes}} & \multicolumn{1}{c|}{\multirow{2}{*}{\textbf{Method}}} & \multirow{2}{*}{\textbf{Accuracy (\%)}} & \multicolumn{4}{c}{\textbf{Communication Burden (B)}}          \\
                                            & \multicolumn{1}{c|}{}                                 &                                         & \textbf{Device} & \textbf{Edge} & \textbf{Cloud} & \textbf{Total} \\ \cline{2-7} 
                                            & EndServe                                              & 89.40                                   & -            & -             & -              & -              \\
                                            & EdgeServe                                             & 90.34                                   & 120.86K      & 120.86K       & -              & 241.72K        \\
                                            & CloudServe                                            & 91.84                                   & 120.86K      & -             & 120.86K        & 241.72K        \\
                                            & ColServe($\alpha=0.2$)                                    & 89.68                                   & 21.62K       & 25.23K        & 3.61K          & 50.46K         \\
                                            & CasServe(setting 3)                         & 90.71                                   & 13.30K       & 14.27K        & 0.96K          & 28.53K         \\
                                            & RecServe($\beta=0.1$)                                    & 90.62                                   & 11.30K       & 13.05K        & 1.75K          & 26.10K         \\
                                            & ColServe($\alpha=0.5$)                                    & 89.77                                   & 63.02K       & 93.45K        & 30.43K         & 186.90K        \\
                                            & CasServe(setting 4)                     & 91.18                                   & 47.94K       & 57.69K        & 9.75K          & 115.38K        \\
                                            & RecServe($\beta=0.3$)                                    & 91.46                                   & 41.34K       & 53.53K        & 12.19K         & 107.06K        \\ \hline
\multirow{11}{*}{\textbf{Yelp Polarity}}   & \multicolumn{1}{c|}{\multirow{2}{*}{\textbf{Method}}} & \multirow{2}{*}{\textbf{Accuracy (\%)}} & \multicolumn{4}{c}{\textbf{Communication Burden (B)}}          \\
                                            & \multicolumn{1}{c|}{}                                 &                                         & \textbf{Device} & \textbf{Edge} & \textbf{Cloud} & \textbf{Total} \\ \cline{2-7} 
                                            & EndServe                                              & 92.46                                   & -            & -             & -              & -              \\
                                            & EdgeServe                                             & 95.28                                   & 25.90M       & 25.90M        & -              & 51.80M         \\
                                            & CloudServe                                            & 96.54                                   & 25.90M       & -             & 25.90M         & 51.80M         \\
                                            & ColServe($\alpha=0.2$)                                    & 92.91                                   & 5.04M        & 6.03M         & 0.99M          & 12.06M         \\
                                            & CasServe(setting 5)                      & 94.58                                   & 2.80M        & 3.62M         & 0.81M          & 7.23M          \\
                                            & RecServe($\beta=0.1$)                                    & 94.63                                   & 3.21M        & 3.55M         & 0.34M          & 7.10M          \\
                                            & ColServe($\alpha=0.5$)                                    & 93.89                                   & 13.01M       & 19.55M        & 6.54M          & 39.10M         \\
                                            & CasServe(setting 6)                      & 95.62                                   & 11.69M       & 13.93M        & 2.24M          & 27.86M         \\
                                            & RecServe($\beta=0.3$)                                    & 96.06                                   & 9.68M        & 12.39M        & 2.71M          & 24.78M         \\ \hline
\multirow{11}{*}{\textbf{Amazon Polarity}} & \multicolumn{1}{c|}{\multirow{2}{*}{\textbf{Method}}} & \multirow{2}{*}{\textbf{Accuracy (\%)}} & \multicolumn{4}{c}{\textbf{Communication Burden (B)}}          \\
                                            & \multicolumn{1}{c|}{}                                 &                                         & \textbf{Device} & \textbf{Edge} & \textbf{Cloud} & \textbf{Total} \\ \cline{2-7} 
                                            & EndServe                                              & 89.66                                   & -            & -             & -              & -              \\
                                            & EdgeServe                                             & 92.88                                   & 154.19M      & 154.19M       & -              & 308.38M        \\
                                            & CloudServe                                            & 95.11                                   & 154.19M      & -             & 154.19M        & 308.38M        \\
                                            & ColServe($\alpha=0.2$)                                    & 90.20                                   & 30.76M       & 36.88M        & 6.12M          & 73.76M         \\
                                            & CasServe(setting 7)                          & 91.71                                   & 18.72M       & 21.04M        & 2.33M          & 42.09M         \\
                                            & RecServe($\beta=0.1$)                                    & 91.88                                   & 17.19M       & 19.09M        & 1.90M          & 38.18M         \\
                                            & ColServe($\alpha=0.5$)                                    & 91.38                                   & 77.21M       & 115.86M       & 38.65M         & 231.72M        \\
                                            & CasServe(setting 6)                      & 93.42                                   & 56.75M       & 72.27M        & 15.52M         & 144.54M        \\
                                            & RecServe($\beta=0.3$)                                    & 94.20                                   & 51.64M       & 68.19M        & 16.56M         & 136.39M        \\ \hline
\end{tabular}
\end{adjustbox}
\end{table*}

\begin{table*}[t]
        \centering
        \caption{BLEU vs communication burden in Seq2Seq tasks.}
        \label{seq2seq} 
        \setlength{\tabcolsep}{10pt}
        \begin{tabular}{c|l|c|cccc}
\hline
\multirow{11}{*}{\textbf{WMT16}}   & \multicolumn{1}{c|}{\multirow{2}{*}{\textbf{Method}}} & \multirow{2}{*}{\textbf{BLEU (\%)}} & \multicolumn{4}{c}{\textbf{Communication Burden   (B)}}        \\
                                   & \multicolumn{1}{c|}{}                                 &                                     & \textbf{Device} & \textbf{Edge} & \textbf{Cloud} & \textbf{Total} \\ \cline{2-7} 
                                   & EndServe                                              & 23.18                               & -            & -             & -              & -              \\
                                   & EdgeServe                                             & 28.87                               & 689.87K      & 689.87K       & -              & 1379.74K       \\
                                   & CloudServe                                            & 29.26                               & 727.11K      & -             & 727.11K        & 1454.22K       \\
                                   & ColServe($\alpha=0.5$)                                    & 26.44                               & 356.49K      & 545.08K       & 188.59K        & 1090.15K       \\
                                   & CasServe(setting 8)                     & 26.00                               & 442.04K      & 674.09K       & 232.04K        & 1348.17K       \\
                                   & RecServe($\beta=0.5$)                                    & 26.60                               & 309.67K      & 454.55K       & 144.88K        & 909.10K        \\
                                   & ColServe($\alpha=0.3$)                                    & 25.00                               & 205.70K      & 269.89K       & 64.19K         & 539.78K        \\
                                   & CasServe(setting 9)                   & 24.30                               & 199.91K      & 258.58K       & 58.67K         & 517.16K        \\
                                   & RecServe($\beta=0.3$)                                    & 25.16                               & 167.60K      & 214.57K       & 46.97K         & 429.14K        \\ \hline
\multirow{11}{*}{\textbf{WMT19}}   & \multicolumn{1}{c|}{\multirow{2}{*}{\textbf{Method}}} & \multirow{2}{*}{\textbf{BLEU (\%)}} & \multicolumn{4}{c}{\textbf{Communication Burden (B)}}          \\
                                   & \multicolumn{1}{c|}{}                                 &                                     & \textbf{Device} & \textbf{Edge} & \textbf{Cloud} & \textbf{Total} \\ \cline{2-7} 
                                   & EndServe                                              & 24.77                               & -            & -             & -              & -              \\
                                   & EdgeServe                                             & 31.28                               & 702.66K      & 702.66K       & -              & 1405.32K       \\
                                   & CloudServe                                            & 31.37                               & 731.13K      & -             & 731.13K        & 1462.26K       \\
                                   & ColServe($\alpha=0.5$)                                    & 28.01                               & 359.49K      & 546.55K       & 187.06K        & 1093.10K       \\
                                   & CasServe(setting 10)                    & 27.54                               & 433.04K      & 660.52K       & 227.48K        & 1321.04K       \\
                                   & RecServe($\beta=0.5$)                                    & 28.38                               & 332.13K      & 488.38K       & 156.25K        & 976.76K        \\
                                   & ColServe($\alpha=0.3$)                                    & 26.52                               & 211.43K      & 278.63K       & 67.20K         & 557.26K        \\
                                   & CasServe(setting 8)                   & 25.75                               & 188.79K      & 251.65K       & 62.86K         & 503.30K        \\
                                   & RecServe($\beta=0.3$)                                    & 26.85                               & 177.97K      & 224.53K       & 46.56K         & 449.06K        \\ \hline
\multirow{11}{*}{\textbf{OPUS100}} & \multicolumn{1}{c|}{\multirow{2}{*}{\textbf{Method}}} & \multirow{2}{*}{\textbf{BLEU (\%)}} & \multicolumn{4}{c}{\textbf{Communication Burden (B)}}          \\
                                   & \multicolumn{1}{c|}{}                                 &                                     & \textbf{Device} & \textbf{Edge} & \textbf{Cloud} & \textbf{Total} \\ \cline{2-7} 
                                   & EndServe                                              & 15.70                               & -            & -             & -              & -              \\
                                   & EdgeServe                                             & 21.44                               & 293.71K      & 293.71K       & -              & 587.42K        \\
                                   & CloudServe                                            & 19.44                               & 388.19K      & -             & 388.19K        & 776.38K        \\
                                   & ColServe($\alpha=0.5$)                                    & 18.40                               & 178.09K      & 278.96K       & 100.87K        & 557.92K        \\
                                   & CasServe(setting 11)                      & 17.84                               & 216.83K      & 364.33K       & 147.50K        & 728.66K        \\
                                   & RecServe($\beta=0.5$)                                    & 18.63                               & 152.67K      & 237.16K       & 84.49K         & 474.32K        \\
                                   & ColServe($\alpha=0.3$)                                    & 16.98                               & 90.56K       & 127.31K       & 36.75K         & 254.62K        \\
                                   & CasServe(setting 12)                     & 17.10                               & 112.45K      & 178.93K       & 66.48K         & 357.86K        \\
                                   & RecServe($\beta=0.3$)                                    & 17.67                               & 66.98K       & 86.53K        & 19.55K         & 173.06K        \\ \hline
\end{tabular}
\label{seq2seq-exp}
\end{table*}

\section{Experiments}
This section validates the performance of RecServe through a series of extensive experiments. We will detail the comprehensive experimental setup, followed by a presentation and analysis of the comparative results.

\subsection{Experimental Setup}
\subsubsection{Tasks and Datasets}
To evaluate RecServe's performance across diverse NLP applications, we consider two distinct task categories with different datasets:
\begin{itemize}
    \item Sequence-to-Class (Seq2Class). We evaluate on five widely-acknowledged benchmark datasets: IMDB \cite{maas2011learning}, SST-2 \cite{socher2013recursive}, Rotten Tomatoes \cite{pang2005seeing}, Yelp Polarity \cite{zhang2015character}, and Amazon Polarity \cite{zhang2015character}, which span diverse textual domains such as movie reviews and product feedback.
    \item Sequence-to-Sequence (Seq2Seq). We select German-to-English (De-En) translation benchmarks from WMT16 \cite{bojar2016findings}, WMT19 \cite{barrault2019findings}, and OPUS100 \cite{tiedemann2012parallel}, which encompass a diverse range of linguistic contexts and complexities.
\end{itemize}

\subsubsection{Hierarchical LLM Deployment}
To simulate realistic multi-tier network environments, we establish a three-tier serving hierarchy comprising device, edge, and cloud nodes. At each computational tier, we deploy LLMs with capacity and complexity appropriate to the node's computational resources. For Seq2Class tasks, we deploy DistilRoBERTa \cite{Sanh2019DistilBERTAD,seq2class-end}, RoBERTa-Base \cite{liu2019roberta,seq2class-edge}, RoBERTa-Large \cite{seq2class-cloud} on end, edge and cloud nodes, respectively. 
For Seq2Seq tasks, we deploy T5-Small \cite{raffel2020exploring,seq2seq-end}, opus-mt \cite{tiedemann2020opus,seq2seq-edge} and fairseq-based wmt19 transformer \cite{kasai2020deep,seq2seq-cloud} at the corresponding tiers. All LLMs are fine-tuned following the official guidelines provided in their respective repositories, with pre-trained weights downloaded from the Huggingface platform.

\subsubsection{Baselines}
We compare RecServe against several representative inference serving paradigms or baseline systems:
\begin{itemize}
    \item Local inference on end devices (EndServe). Tasks are processed entirely on end devices without any offloading to remote servers \cite{liu2024mobilellm,hu2024minicpm}.
    \item Full offloading to edge server (EdgeServe). Tasks are fully offloaded to the edge server for inference \cite{padmanabhan2023gemel}.
    \item Full offloading to cloud data center (CloudServe). Tasks are fully offloaded to a centralized cloud data center for inference \cite{fu2024serverlessllm,zhong2024distserve}.
    \item Multi-tier collaborative serving with quality-independent partial offloading (ColServe). Tasks are partially offloaded to higher-tier servers without considering inference quality. In this paper, we consider a fixed probability $\alpha$ to control the offloading process.
    \item Multi-tier collaborative serving with model cascades (CasServe). Tasks are sequentially processed using hierarchical model cascades deployed across multiple tiers. Offloading to a higher-tier server occurs only if the inference confidence at the current tier does not meet predefined thresholds \cite{lebovitz2023efficient}. In this paper, we define thresholds $t_{end}$ and $t_{edge}$ for end devices and edge nodes, respectively.
\end{itemize}

\subsubsection{Hyper-parameters}
We outline the hyperparameter configurations for both the considered baselines and our proposed RecServe, aiming to ensure fair and valid comparisons across tasks and datasets.
\begin{itemize}
\item For EndServe, EdgeServe, and CloudServe, no additional hyperparameters are required.
\item For ColServe, we consider the cases where $\alpha \in \{0.2,0.3,0.5\}$.
\item For CasServe, we apply careful tuning of $t_{end}$ and $t_{edge}$ to achieve comparability with relevant baselines. Specifically, twelve different hyper-parameter settings (setting 1-12) are considered with threshold combinations $(t_{end}, t_{edge})$ as follows: (0.85, 0.6), (0.99, 0.7), (0.97, 0.7), (0.998, 0.95), (0.85, 0.75), (0.995, 0.8), (0.9, 0.6), (0.05, 0.001), (0.025, 0.0002), (0.055, 0.001), (0.07, 0.01), and (0.03, 0.001).
\item For our proposed RecServe, we set $k=10000$, and consider the cases where $\beta \in \{0.1,0.3,0.5\}$.
\end{itemize}

\subsubsection{Evaluation Metrics}
We evaluate the performance of serving systems based on the following metrics:
\begin{itemize}
    \item Precision. We measure inference accuracy for Seq2Class tasks and BLEU scores for Seq2Seq tasks to quantify predictive performance.
    \item Communication burden. We quantify data transmission volume per computing node and aggregate this across all nodes to assess system-wide communication efficiency.
    \item Precision-communication trade-off. We analyze the relationship between precision and communication burden through comparative performance plots.
\end{itemize}

\begin{figure*}[t]
    \centering
    \begin{minipage}{\linewidth}
        \centering
        \includegraphics[width=\textwidth]{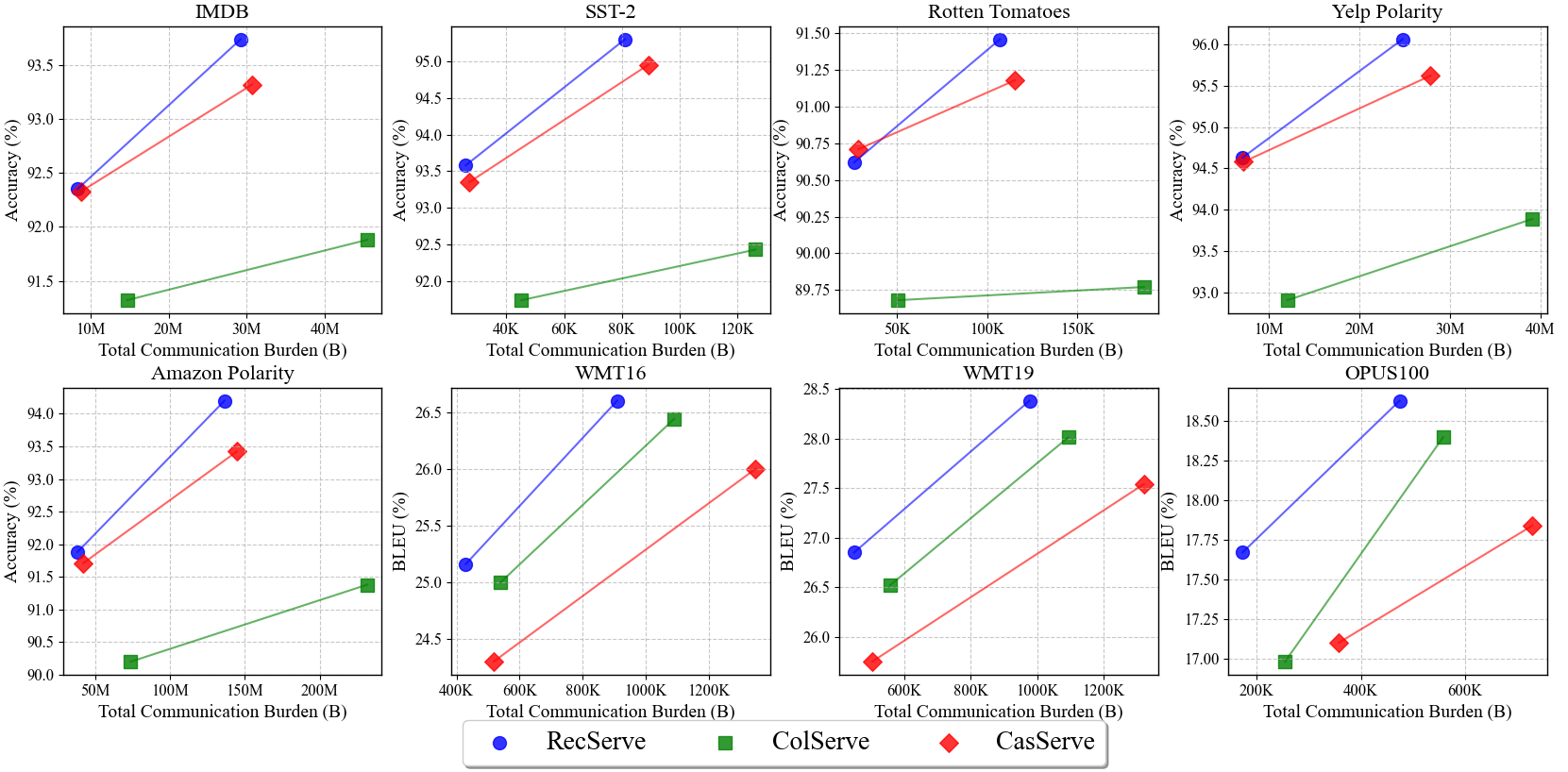}
        \caption{Visualization of precision vs communication burden for multi-tier serving methods across eight datasets.}
        \label{vis-serve-cmp}
    \end{minipage}\par\medskip
\end{figure*}

\begin{figure}[t]
    \centering
    \begin{minipage}{\linewidth}
        \centering
        \includegraphics[width=0.9\textwidth]{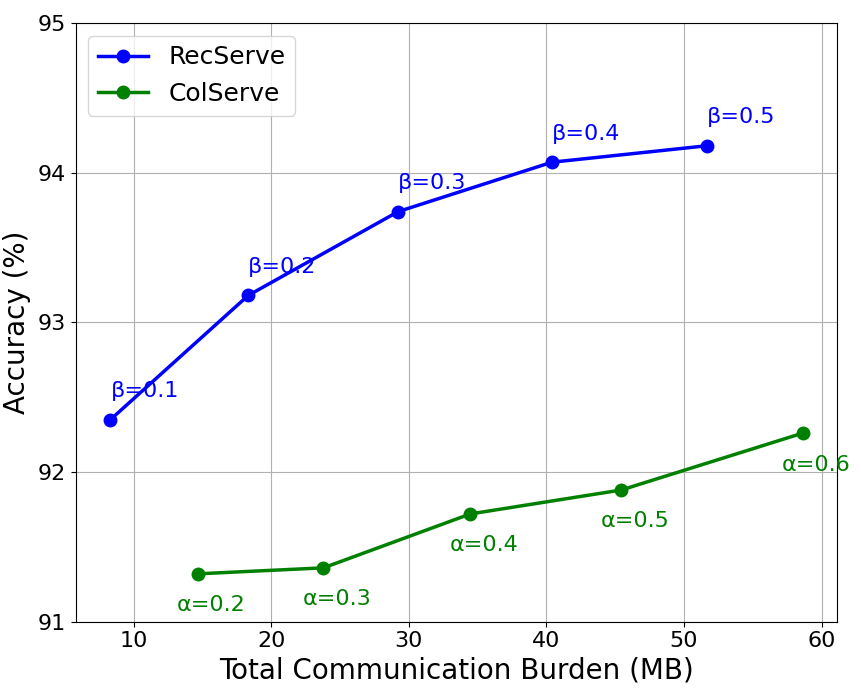}
        \caption{Comparison of RecServe and ColServe with different offload configurations.}
        \label{ablation-1}
    \end{minipage}\par\medskip
\end{figure}

\subsection{Experimental Results}
\subsubsection{Evaluation on Seq2Class Tasks}
TABLE \ref{seq2class-exp} reports the inference accuracy and communication burden of various methods across five datasets: IMDB, SST-2, Rotten Tomatoes, Yelp Polarity, and Amazon Polarity. As displayed, CloudServe achieves the highest accuracy across all datasets, with values of 94.25\% (IMDB), 95.76\% (SST-2), 91.84\% (Rotten Tomatoes), 96.54\% (Yelp Polarity), and 95.11\% (Amazon Polarity). However, this comes at the cost of substantial total communication burden, with 60.82MB, 179.60KB, 241.72KB, 51.80MB, and 308.38MB, on these five datasets, respectively. EdgeServe offers slightly lower accuracy with no improvement in communication burden compared to CloudServe.
In contrast, our proposed RecServe ($\beta=0.3$) delivers competitive accuracies compared to CloudServe and EdgeServe, which are 93.74\% on IMDB, 95.30\% on SST-2, 91.46\% on Rotten Tomatoes, 96.06\% on Yelp Polarity, and 94.20\% on Amazon Polarity. At the same time, the total communication burden of RecServe ($\beta=0.3$) is significantly reduced by over 50\% compared to CloudServe and EdgeServe, which are only 29.22MB on IMDB, 81.12KB on SST-2, 107.06KB on Rotten Tomatoes, 24.78MB on Yelp Polarity, and 136.39MB on Amazon Polarity, respectively. In addition, RecServe significantly improves accuracy across all datasets compared to EndServe.
Against multi-tier baselines such as ColServe and CasServe, RecServe demonstrates superior performance in both accuracy and communication efficiency. On the IMDB dataset, for instance, ColServe ($\alpha=0.5$) achieves 91.88\% accuracy with 45.44MB total communication burden, while CasServe (setting 2) yields 93.32\% with 30.70MB. RecServe ($\beta=0.3$), by comparison, achieves 93.74\% accuracy with only 29.22MB total communication burden. Similar patterns are observed across other datasets, with RecServe outperforming these multi-tier serving methods in most configurations. A visualization of these trade-offs is provided in subsection \ref{vis-section}.


\subsubsection{Evaluation on Seq2Seq Tasks}
TABLE \ref{seq2seq-exp} summarizes the BLEU scores and communication burden over WMT16, WMT19, and OPUS100 datasets. As illustrated, CloudServe achieves the highest BLEU scores in most cases, but at the cost of the largest communication burden. EdgeServe, while slightly more efficient in communication, still incurs considerable overhead.
On the contrary, RecServe ($\beta=0.5$) offers a strong balance between serving performance and communication efficiency, achieving BLEU scores comparable to CloudServe while reducing communication burden by over 33\%. At lower values of $\beta$ (e.g., $\beta=0.3$), RecServe further reduces communication burden by more than 50\%, still maintaining BLEU scores that significantly outperform EndServe. 
Compared to ColServe and CasServe, RecServe consistently delivers higher BLEU scores with lower communication burden. For example, on WMT19, ColServe ($\alpha=0.3$) achieves a BLEU score of 26.52\% with a 557.26KB communication burden, and CasServe (setting 8) obtains 25.75\% with 503.30KB. In contrast, RecServe ($\beta=0.3$) achieves 26.85\% with only 449.06KB. Further precision-communication comparisons of these multi-tier serving methods are elaborated in subsection \ref{vis-section}.

\begin{figure}[t]
    \centering
    \begin{minipage}{\linewidth}
        \centering
        \includegraphics[width=\textwidth]{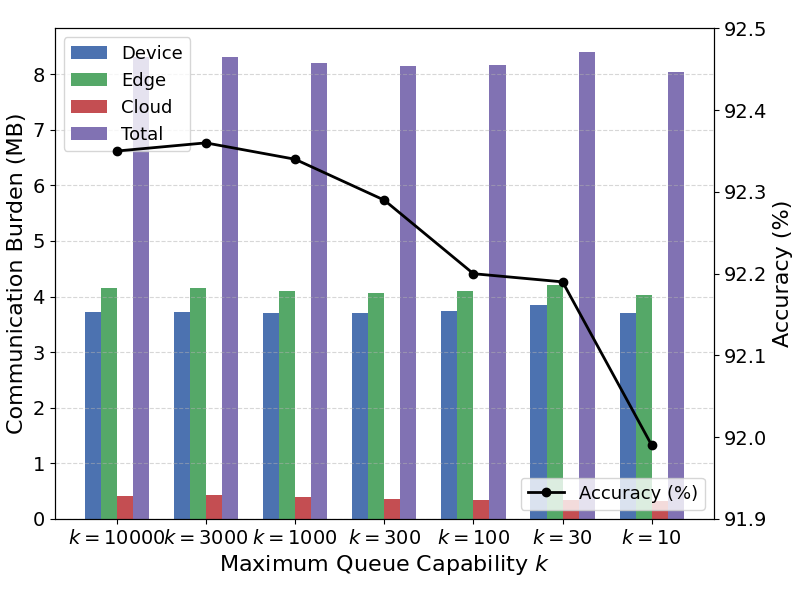}
        \caption{Effect of maximum queue capability on RecServe’s inference accuracy ($\beta=0.1$) and communication burden.}
        \label{ablation-2}
    \end{minipage}\par\medskip
\end{figure}

\subsubsection{Visualization of Precision-Communication Trade-off}
\label{vis-section}
Fig. \ref{vis-serve-cmp} presents the comparison of the performance of multi-tier serving methods (ColServe, CasServe, and RecServe) across eight datasets, evaluating both precision (measured by accuracy or BLEU score) and communication burden.
Notably, RecServe (represented by the brown dots and lines) is positioned in the upper left relative to both ColServe (red) and CasServe (purple) in nearly all cases, indicating higher accuracy achieved at lower communication burden. For instance, on the IMDB dataset, RecServe achieves the highest accuracy at all levels of communication burden, with its performance reaching up to 93.5\% with less than 30MB total communication burden. In contrast, ColServe and CasServe start with lower accuracy, and their accuracy improvements are slower as the communication burden increases.
RecServe continues to excel in tasks involving BLEU scores. For example, in the WMT16 dataset, RecServe maintains a significant lead in BLEU scores, surpassing 26.5\% at total communication burdens below 1000KB. Meanwhile, both CasServe and ColServe consistently fall short of this threshold.

\begin{figure}[t]
    \centering
    \begin{minipage}{\linewidth}
        \centering
        \includegraphics[width=1.0\textwidth]{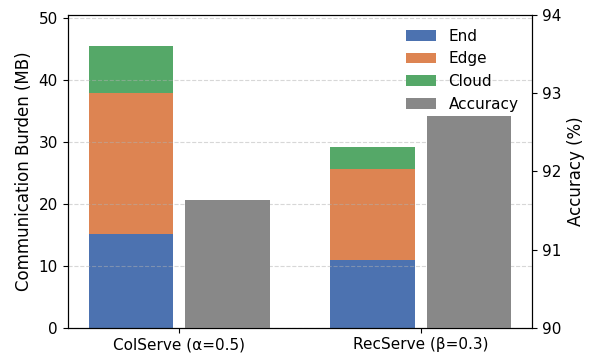}
        \caption{Comparison of RecServe and ColServe with DeBERTa-large deployed on the cloud. For each method, the left bar shows the stacked communication burden across end, edge, and cloud tiers, while the right bar indicates the corresponding accuracy.}
        \label{ablation-model}
    \end{minipage}\par\medskip
\end{figure}

\section{Ablation Study}
This section further investigates the performance characteristics of RecServe by examining the impact of key parameters. We will illustrate on the ablation studies focusing on three aspects: the impact of the dynamic offloading threshold quantile $\beta$, the maximum queue capability $k$, and the choice of cloud-side LLM. All the experimental results are conducted on the IMDB dataset, following the same experimental configuration as in the main experiments.

\subsection{Impact of Dynamic Offloading Threshold Quantile}
We first investigate how the offloading threshold quantile $\beta$ affects the trade-off between inference accuracy and total communication burden. Fig. \ref{ablation-1} illustrates the performance of RecServe with different values of $\beta$ (ranging from 0.1 to 0.5), in comparison with ColServe under varying offloading probabilities $\alpha$. As shown in the figure, increasing $\beta$ leads to higher accuracy but also increases the total communication burden. Specifically, when $\beta$ is set to 0.1, RecServe achieves an accuracy of 92.3\% with a communication burden less than 10 MB. As $\beta$ increases to 0.5, accuracy rises to 94.2\% while the communication burden increases to more than 50 MB.
Comparing RecServe with ColServe, there exists at least one RecServe configuration that simultaneously achieves higher accuracy and lower communication burden in each configuration evaluated for ColServe. For example, at a communication burden of around 30 MB, RecServe with $\beta=0.3$ attains an accuracy of about 93.7\%, substantially outperforming ColServe with $\alpha=0.5$, which only reaches less than 91.9\%. These results demonstrate the effectiveness of RecServe's recursive offloading strategy in achieving a superior trade-off between communication efficiency and serving performance.

\subsection{Impact of Maximum Queue Capability}
We further analyze the influence of the maximum queue capability $k$ in the historical confidence queue on both communication burden and inference accuracy. Fig. \ref{ablation-2} illustrates the results of RecServe as $k$ varies from 10 to 10000. The bar charts show the communication burden at each network tier (device, edge, cloud), as well as the total burden, while the black line represents the corresponding inference accuracy.
As $k$ increases, the communication burden remains largely stable, while accuracy initially improves before reaching a plateau beyond $k=300$. For larger values of $k$ (e.g., $k=1000$ or higher), the accuracy stabilizes at approximately 92.35\%, with little further change in communication burden.
This behavior can be attributed to the stability of the dynamic threshold estimation: a larger queue provides a more reliable and representative history of LLM confidence, enabling more accurate offloading decisions. However, excessively large $k$ yields minimal additional benefits, as the confidence distribution converges. Therefore, in practical deployments, setting $k$ to a moderate value (e.g., 300–1000) is sufficient to achieve near-optimal performance without unnecessary storage or computational overhead.

\begin{figure}[t]
    \centering
    \begin{minipage}{\linewidth}
        \centering
        \includegraphics[width=\textwidth]{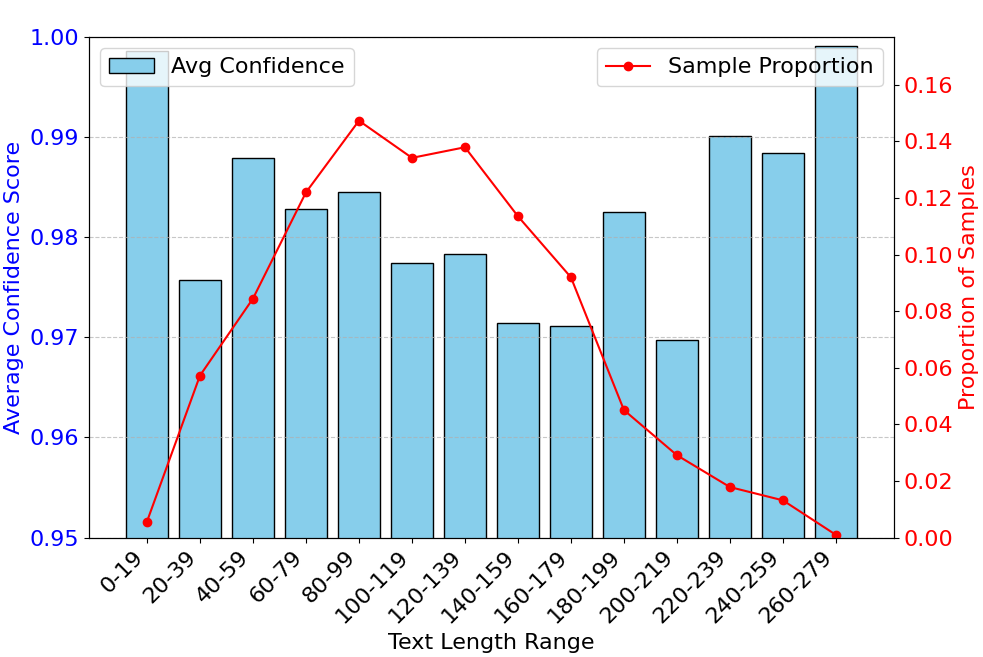}
        \caption{Confidence score and sample distribution across text lengths. Results are derived from T5-Small on the Rotten Tomatoes dataset.}
        \label{length-confidence}
    \end{minipage}\par\medskip
\end{figure}

\subsection{Impact of Cloud-side LLM}
\textcolor{black}{To evaluate the robustness of RecServe against variations in the cloud-side LLM, we replace the original cloud-tier model with a version of DeBERTa-Large fine-tuned on SST-2 \cite{he2020deberta,ablation-model-deberta} without any tuning for multi-tier coordination. As shown in Fig. \ref{ablation-model}, RecServe achieves a substantial improvement in serving performance, whose accuracy is approximately 1\% higher than ColServe. In terms of communication, RecServe still maintains superior efficiency. The total communication burden of RecServe is less than 30MB, significantly lower than ColServe's approximately 45MB, representing a reduction of over 30\%. These results demonstrate that RecServe is robust to the choice of the cloud-side LLM, confirming its effectiveness across different model configurations in practical deployment scenarios.}

\begin{figure}[t]
    \centering
    \begin{minipage}{\linewidth}
        \centering
        \includegraphics[width=0.95\textwidth]{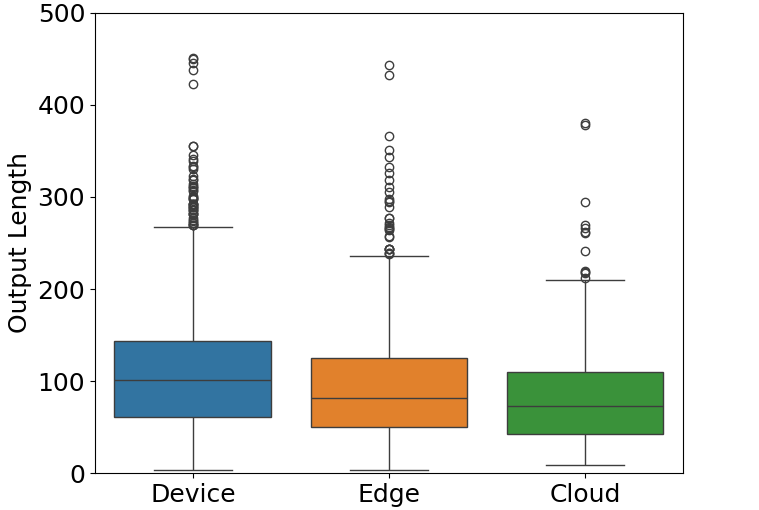}
        \caption{Output length distribution across device, edge, and cloud tiers. Results are derived from WMT16 dataset, taking $\beta=0.5$.}
        \label{dist-length}
    \end{minipage}\par\medskip
\end{figure}

\section{Discussion}
This section extends the analysis of RecServe by considering its wider implications and inherent challenges. We will elaborate on the broader impact of the proposed framework, investigate the reasons behind deviations from our theoretical communication models, and discuss limitations and how they can be mitigated in real-world deployments.

\subsection{Broader Impact}
\color{black}
\subsubsection{Cross-tier Load Balancing}
RecServe achieves cross-tier load balancing through intelligent orchestration of inference tasks across the heterogeneous device-edge-cloud architecture, optimizing resource utilization and preventing computational bottlenecks at any single tier.
By enabling end devices to handle simple inference tasks locally using lightweight LLMs, RecServe prevents resource-constrained nodes from becoming overwhelmed while maximizing their contribution to overall system throughput. At the same time, the tiered offload design of RecServe prevents the premature escalation of all but the most demanding tasks to the cloud, ensuring that edge network capabilities are fully leveraged for moderately complex computations.
\color{black}

\subsubsection{End-to-End Service Latency}
RecServe offers potential benefits in end-to-end service latency reduction for the majority of inference tasks. By intelligently routing dominant simple tasks to smaller LLMs at lower tiers, RecServe simultaneously avoids the higher inference times of larger LLMs and eliminates the substantial round-trip delays associated with device-cloud communications. Furthermore, RecServe's distributed inference architecture naturally load-balances inference tasks across the network hierarchy, reducing congestion at higher tiers during peak periods when centralized services often face exponential increases in queuing delays. These latency benefits are particularly valuable for interactive applications where perceived responsiveness directly impacts user engagement and satisfaction.


\subsection{Discrepancies Between Theoretical and Actual Communication Burdens}
\label{comm-bias}
While our theoretical analysis offers an approximation of communication burden under idealized assumptions, the actual results differ somewhat in practice due to several factors.

\begin{itemize}
\item Cold start of historical confidence queues. The theoretical analysis presumes a historical confidence queue with sufficient samples to estimate dynamic offloading thresholds accurately. However, during system initialization or early-stage deployment, the queues may lack representative samples, leading to suboptimal threshold estimation and irregular offloading decisions.

\item Correlation between input length and confidence. The theoretical analysis assumes independence between input sequence length $|x|$ and the output confidence score $C_{M_i, \tau}(x)$, which is impractical in reality, as shown in Fig. \ref{length-confidence}.

\item Variation in output length distribution across tiers. The theoretical framework assumes uniform output length $|y|$ across LLMs deployed on all tiers. However, actual output lengths may differ due to variations in LLM capacity, decoding strategies, or sample-specific generation behavior, as illustrated in Fig. \ref{dist-length}.
\end{itemize}
These aforementioned factors collectively explain the observed deviations of the actual communication burden from that predicted by our theoretical analysis.

\color{black}
\subsection{Limitations and Countermeasures}
\subsubsection{Upper-Tier Node Unavailability} 
Current design of RecServe assumes that upper-tier nodes (e.g., edge servers and cloud data centers) are continuously available to process offloaded tasks, as listed in Eq. (\ref{recur-offload}). However, in real-world deployments, these nodes may occasionally become temporarily unavailable due to factors such as resource overload, network congestion, or scheduled maintenance.
To address this limitation, we define a function $A(M')$ that evaluates the availability of the next-tier node $M'$ for task offloading. If offload is advised ($C_{M,\tau}(x) < T_{M,\tau}(\beta)$ and $M \ne M_n$) but the target node $M'$ is unavailable ($\neg A(M')$), than the current node $M$ should cease further upward recursion for that specific task and shoulder the responsibility for final task execution.
Incorporating this availability check, the unavailability-tolerant recursive offloading decision policy $\mathcal{D}_{ut}$ can be formulated as:
\begin{equation}
\begin{array}{l}
\mathcal{D}_{ut}(x, M, \tau) =\\
\begin{cases}
M,  \text{if } C_{M, \tau}(x) \geq T_{M, \tau}(\beta) \text{ or } \neg A(M') \\
\mathcal{D}_{ut}(x, M', \tau),\\ \text{if } M \neq M_n \text{ and } C_{M, \tau}(x) < T_{M, \tau}(\beta) \text{ and } A(M'),
\end{cases}
\end{array}
\end{equation}
Accordingly, the revised inference output is given by the following equations:
\begin{equation}
y = M^{**}(x), M^{**}=\mathcal{D}_{ut}(x, M_1, \tau).
\end{equation}
This fault-tolerant mechanism modification allows RecServe to gracefully degrade service by completing the task at the current available tier when an intended offload to a higher-tier node is prevented by that node's unavailability, ensuring that RecServe maintains operational resilience in the face of unpredictable network or node failures within the higher tiers.

\subsubsection{Inference Serving Under Communication Budget}
As discussed in subsection \ref{comm-bias}, we observe the systematic bias between the theoretical expected communication burden (denoted as $E_{theo}[\text{Comm-RecServe}(\beta)]$) and the expected actual communication burden (denoted as $E_{act}[\text{Comm-RecServe}(\beta)]$). This discrepancy complicates selecting an offloading quantile $\beta$ such that the actual burden meets a precise communication budget $B_{\text{comm}}$:
\begin{equation}
    E_{act}[\text{Comm-RecServe}(\beta)] \approx B_{comm}.
\end{equation}
To address this challenge and ensure that RecServe adheres to a predefined communication budget, we propose an online feedback-based calibration mechanism that dynamically adjusts $\beta$ to align the actual communication burden with the target budget. The designed procedure is as follows:

\begin{enumerate}
    \item Select an initial value $\beta_0$ such that the theoretical expected communication burden equals the target budget:
    \begin{equation}
        E_{theo}[\text{Comm-RecServe}(\beta_0)] = B_{comm}.
    \end{equation}
    
    \item Over a window of $R$ requests, measure the actual expected communication burden for the current $\beta_t$, where $t$ indicates the current calibration round.

    \item Compute the feedback ratio $\gamma(\beta_t)$ as follows:
    \begin{equation}
        \gamma(\beta_t) = \frac{E_{act}[\text{Comm-RecServe}(\beta_t)]}{B_{comm}}.
    \end{equation}
    When $\gamma(\beta_t) > 1$, the actual burden exceeds the budget, and vice versa.

    \item Update $\beta$ using a proportional control strategy:
    \begin{equation}
        \beta_{t+1} = \frac{\beta_t}{\gamma(\beta_t)^\eta}
    \end{equation}
    where $\eta$ is a hyperparameter that controls the adjustment rate.

    \item  Repeat steps 2–4 until convergence ($\gamma(\beta_t) \approx 1$), ensuring that the actual expected communication burden closely matches the target budget.
\end{enumerate}
This iterative calibration mechanism enables RecServe to operate within a specified communication budget by adaptively tuning its offloading parameter $\beta$ based on real-time inference feedback.

\color{black}

\section{Related Work}
\subsection{Large Language Model}
The development of LLM is fundamentally rooted in the Transformer architecture \cite{vaswani2017attention}, evolving from foundation model design to generalized task adaptation and large-scale deployment. One of the earliest representative LLMs, BERT \cite{devlin2019bert}, introduced a bidirectional Transformer encoder to effectively capture contextual text information, achieving significant success in natural language understanding tasks. Its successor, RoBERTa \cite{liu2019roberta}, further enhanced performance by increasing the training corpus size, extending the number of training epochs, and removing the next sentence prediction objective.
Since 2020, LLMs have entered the era of hundred-billion-parameter models, which became a mainstream trend by 2022. A key milestone in this evolution was GPT-3.5 \cite{openai2022gpt3-5,brown2020language}, which demonstrated notable capability in natural language generation. GPT-4 series \cite{achiam2023gpt,shahriar2024putting} further achieve superior performance in multimodal understanding and daily tasks. Concurrently, the Claude series \cite{claude3,claude3-5,claude3-7} has distinguished itself through a strong focus on safety and alignment, enhancing the model’s ability to understand and respond to human intent while maintaining robust language comprehension.
In the open-source community, LLaMA series \cite{touvron2023llama,touvron2023llama2,grattafiori2024llama} has been widely adopted in both academia and industry for its efficiency, performance, and flexibility.
Alibaba’s Qwen series \cite{bai2023qwen,wang2024qwen2,bai2025qwen2} focuses on multilingual capabilities and multimodal inference. DeepSeek-R1 \cite{guo2025deepseek}, leveraging a reinforcement learning-driven pretraining paradigm, endows models with powerful reasoning capabilities and achieves outstanding performance in tasks such as mathematics, coding, and logic.

\subsection{Large Language Model Serving}
Efficient inference services for LLMs are essential to unlocking their full potential in real-world applications.
Mainstream LLM serving solutions rely on cloud-centric infrastructure, leveraging abundant computing resources available in cloud data centers to achieve multi-GPU parallel inference. 
PagedAttention \cite{kwon2023efficient} applies a virtual memory-inspired paging mechanism to optimize key-value (KV) cache management, reducing memory waste and improving throughput.
CacheGen \cite{liu2024cachegen} introduces a KV cache compression method for improving bandwidth utilization and dynamically adjusting compression levels to balance loading delay and generation quality. 
ServerlessLLM \cite{fu2024serverlessllm} significantly reduces inference latency in serverless environments through fast multi-tier checkpoint loading, efficient live migration, and optimized model scheduling.
DistServe \cite{zhong2024distserve} separates pre-filling and decoding computations onto different GPUs to improve parallelism and resource allocation.
In edge-based serving scenarios, optimizing resource efficiency becomes a central challenge due to the scarce nature of computing capabilities on edge computing nodes. AdaInf \cite{shubha2023adainf} conducts a data drift-aware scheduling method that combines incremental retraining and service-level-objective-compliant inference with optimized CPU-GPU memory communications. EdgeShard \cite{zhang2024edgeshard} takes advantage of model partition by distributing model shards across multiple edge resource-constrained devices and applying dynamic programming algorithms to optimize inference latency and throughput.
Additionally, edge-cloud collaborative serving offers a promising approach for balancing performance and resource efficiency. PerLLM \cite{yang2024perllm} introduces a constraint-aware personalized scheduling framework in edge-cloud environments to meet latency requirements while reducing energy consumption.

\subsection{Task Offloading for Artificial Intelligence}
Task offloading has long been a research focus in the fields of edge and cloud computing \cite{islam2021survey,saeik2021task}, which offload an application module from the local computing node to a remote server for execution and retrieve the results upon completion. With the rapid development of artificial intelligence (AI), task offloading has been extended to the training and inference processes of both Deep Neural Networks (DNN) and LLMs. Split Computing \cite{matsubara2022split} executes the initial layers of a neural network on a local device and offloads the remaining layers to edge servers, achieving low-latency response with acceptable accuracy. Mohammed \cite{mohammed2020distributed} proposes a DNN offloading strategy based on adaptive fine-grained partitioning and a swap-matching-based algorithm to dynamically divide DNNs into sub-layer components and efficiently distribute inference tasks across the network, significantly reducing inference latency. Mandheling \cite{xu2022mandheling} introduces the first efficient on-device training system by integrating mixed-precision training with on-chip Digital Signal Processing (DSP) offloading, saving both convergence time and energy consumption. In addition, InstInfer \cite{pan2024instinfer} offloads attention computation and KV cache management to Computational Storage Drives (CSDs) during decoding, alleviating PCIe bottlenecks and improving inference efficiency for long-context LLMs.

\color{black}
\section{Conclusion}
This paper proposes RecServe, a recursive offloading framework designed for LLM serving in multi-tier networks, which avoid unnecessary cloud-based LLM invocations while addressing the critical balance between inference quality and communication burden. 
RecServe deploys LLMs with progressively increasing capabilities across device, edge, and cloud nodes. By combining task-specific confidence evaluations with dynamic offloading guided by sliding historical knowledge queues, RecServe enables the majority of inference tasks to be efficiently completed at lower tiers, while selectively offloading only the more complex tasks with insufficient confidence to higher-tier nodes. Theoretical analysis provides parameter bounds that facilitate the alleviation of expected communication burden and computational costs, respectively. Extensive experiments demonstrate that RecServe significantly outperforms existing multi-tier LLM serving approaches in both service quality and communication efficiency, and achieves substantial reductions in communication burden compared to cloud-only deployments.
\color{black}

\section*{Acknowledgement}
We gratefully acknowledge Zixuan Li (Institute of Automation, Chinese Academy of Sciences), Yanyu Ren (Tsinghua University), and Yan Lei (Institute of Computing Technology, Chinese Academy of Sciences)  for their technical guideline.

\ifCLASSOPTIONcaptionsoff
  \newpage
\fi

\bibliographystyle{IEEEtran}

\begin{thebibliography}{10}
\providecommand{\url}[1]{#1}
\csname url@samestyle\endcsname
\providecommand{\newblock}{\relax}
\providecommand{\bibinfo}[2]{#2}
\providecommand{\BIBentrySTDinterwordspacing}{\spaceskip=0pt\relax}
\providecommand{\BIBentryALTinterwordstretchfactor}{4}
\providecommand{\BIBentryALTinterwordspacing}{\spaceskip=\fontdimen2\font plus
\BIBentryALTinterwordstretchfactor\fontdimen3\font minus
  \fontdimen4\font\relax}
\providecommand{\BIBforeignlanguage}[2]{{%
\expandafter\ifx\csname l@#1\endcsname\relax
\typeout{** WARNING: IEEEtran.bst: No hyphenation pattern has been}%
\typeout{** loaded for the language `#1'. Using the pattern for}%
\typeout{** the default language instead.}%
\else
\language=\csname l@#1\endcsname
\fi
#2}}
\providecommand{\BIBdecl}{\relax}
\BIBdecl

\bibitem{xu2021edge}
D.~Xu, T.~Li, Y.~Li, X.~Su, S.~Tarkoma, T.~Jiang, J.~Crowcroft, and P.~Hui,
  ``Edge intelligence: Empowering intelligence to the edge of network,''
  \emph{Proceedings of the IEEE}, vol. 109, no.~11, pp. 1778--1837, 2021.

\bibitem{yang2022edge}
B.~Yang, B.~Wu, Y.~You, C.~Guo, L.~Qiao, and Z.~Lv, ``Edge intelligence based
  digital twins for internet of autonomous unmanned vehicles,'' \emph{Software:
  Practice and Experience}, 2022.

\bibitem{nasir2022enabling}
M.~Nasir, K.~Muhammad, A.~Ullah, J.~Ahmad, S.~W. Baik, and M.~Sajjad,
  ``Enabling automation and edge intelligence over resource constraint iot
  devices for smart home,'' \emph{Neurocomputing}, vol. 491, pp. 494--506,
  2022.

\bibitem{gong2020edgerec}
Y.~Gong, Z.~Jiang, Y.~Feng, B.~Hu, K.~Zhao, Q.~Liu, and W.~Ou, ``Edgerec:
  recommender system on edge in mobile taobao,'' in \emph{Proceedings of the
  29th ACM International Conference on Information \& Knowledge Management},
  2020, pp. 2477--2484.

\bibitem{antunes2022federated}
R.~S. Antunes, C.~Andr{\'e}~da Costa, A.~K{\"u}derle, I.~A. Yari, and
  B.~Eskofier, ``Federated learning for healthcare: Systematic review and
  architecture proposal,'' \emph{ACM Transactions on Intelligent Systems and
  Technology (TIST)}, vol.~13, no.~4, pp. 1--23, 2022.

\bibitem{liu2023online}
Q.~Liu, S.~Sun, M.~Liu, Y.~Wang, and B.~Gao, ``Online spatio-temporal
  correlation-based federated learning for traffic flow forecasting,''
  \emph{arXiv preprint arXiv:2302.08658}, 2023.

\bibitem{kanagavelu2021federated}
R.~Kanagavelu, Z.~Li, J.~Samsudin, S.~Hussain, F.~Yang, Y.~Yang, R.~S.~M. Goh,
  and M.~Cheah, ``Federated learning for advanced manufacturing based on
  industrial iot data analytics,'' \emph{Implementing Industry 4.0: The Model
  Factory as the Key Enabler for the Future of Manufacturing}, pp. 143--176,
  2021.

\bibitem{mcmahan2017communication}
B.~McMahan, E.~Moore, D.~Ramage, S.~Hampson, and B.~A. y~Arcas,
  ``Communication-efficient learning of deep networks from decentralized
  data,'' in \emph{Artificial intelligence and statistics}.\hskip 1em plus
  0.5em minus 0.4em\relax PMLR, 2017, pp. 1273--1282.

\bibitem{li2020federated}
T.~Li, A.~K. Sahu, M.~Zaheer, M.~Sanjabi, A.~Talwalkar, and V.~Smith,
  ``Federated optimization in heterogeneous networks,'' \emph{Proceedings of
  Machine Learning and Systems}, vol.~2, pp. 429--450, 2020.

\bibitem{kulkarni2020survey}
V.~Kulkarni, M.~Kulkarni, and A.~Pant, ``Survey of personalization techniques
  for federated learning,'' in \emph{2020 Fourth World Conference on Smart
  Trends in Systems, Security and Sustainability (WorldS4)}.\hskip 1em plus
  0.5em minus 0.4em\relax IEEE, 2020, pp. 794--797.

\bibitem{tan2022towards}
A.~Z. Tan, H.~Yu, L.~Cui, and Q.~Yang, ``Towards personalized federated
  learning,'' \emph{IEEE Transactions on Neural Networks and Learning Systems},
  2022.

\bibitem{t2020personalized}
C.~T~Dinh, N.~Tran, and J.~Nguyen, ``Personalized federated learning with
  moreau envelopes,'' \emph{Advances in Neural Information Processing Systems},
  vol.~33, pp. 21\,394--21\,405, 2020.

\bibitem{mills2021multi}
J.~Mills, J.~Hu, and G.~Min, ``Multi-task federated learning for personalised
  deep neural networks in edge computing,'' \emph{IEEE Transactions on Parallel
  and Distributed Systems}, vol.~33, no.~3, pp. 630--641, 2021.

\bibitem{jin2022personalized}
H.~Jin, D.~Bai, D.~Yao, Y.~Dai, L.~Gu, C.~Yu, and L.~Sun, ``Personalized edge
  intelligence via federated self-knowledge distillation,'' \emph{IEEE
  Transactions on Parallel and Distributed Systems}, vol.~34, no.~2, pp.
  567--580, 2022.

\bibitem{wu2022communication}
C.~Wu, F.~Wu, L.~Lyu, Y.~Huang, and X.~Xie, ``Communication-efficient federated
  learning via knowledge distillation,'' \emph{Nature communications}, vol.~13,
  no.~1, pp. 1--8, 2022.

\bibitem{sattler2021cfd}
F.~Sattler, A.~Marban, R.~Rischke, and W.~Samek, ``Cfd: Communication-efficient
  federated distillation via soft-label quantization and delta coding,''
  \emph{IEEE Transactions on Network Science and Engineering}, vol.~9, no.~4,
  pp. 2025--2038, 2021.

\bibitem{zhang2021parameterized}
J.~Zhang, S.~Guo, X.~Ma, H.~Wang, W.~Xu, and F.~Wu, ``Parameterized knowledge
  transfer for personalized federated learning,'' \emph{Advances in Neural
  Information Processing Systems}, vol.~34, pp. 10\,092--10\,104, 2021.

\bibitem{jee2023communication}
Y.~J. Cho, J.~Wang, T.~Chirvolu, and G.~Joshi, ``Communication-efficient and
  model-heterogeneous personalized federated learning via clustered knowledge
  transfer,'' \emph{IEEE Journal of Selected Topics in Signal Processing},
  vol.~17, no.~1, pp. 234--247, 2023.

\bibitem{wu2023fedict}
Z.~Wu, S.~Sun, Y.~Wang, M.~Liu, X.~Jiang, and B.~Gao, ``Fedict: Federated
  multi-task distillation for multi-access edge computing,'' \emph{arXiv
  preprint arXiv:2301.00389}, 2023.

\bibitem{zhu2021dataicml}
\BIBentryALTinterwordspacing
Z.~Zhu, J.~Hong, and J.~Zhou, ``Data-free knowledge distillation for
  heterogeneous federated learning,'' in \emph{Proceedings of the 38th
  International Conference on Machine Learning, {ICML} 2021, 18-24 July 2021,
  Virtual Event}, ser. Proceedings of Machine Learning Research, M.~Meila and
  T.~Zhang, Eds., vol. 139.\hskip 1em plus 0.5em minus 0.4em\relax {PMLR},
  2021, pp. 12\,878--12\,889. [Online]. Available:
  \url{http://proceedings.mlr.press/v139/zhu21b.html}
\BIBentrySTDinterwordspacing

\bibitem{jeong2018communication}
E.~Jeong, S.~Oh, H.~Kim, J.~Park, M.~Bennis, and S.-L. Kim,
  ``Communication-efficient on-device machine learning: Federated distillation
  and augmentation under non-iid private data,'' \emph{arXiv preprint
  arXiv:1811.11479}, 2018.

\bibitem{malkov2018efficient}
Y.~A. Malkov and D.~A. Yashunin, ``Efficient and robust approximate nearest
  neighbor search using hierarchical navigable small world graphs,'' \emph{IEEE
  transactions on pattern analysis and machine intelligence}, vol.~42, no.~4,
  pp. 824--836, 2018.

\bibitem{li2019fedmd}
D.~Li and J.~Wang, ``Fedmd: Heterogenous federated learning via model
  distillation,'' \emph{arXiv preprint arXiv:1910.03581}, 2019.

\bibitem{lecun1998gradient}
Y.~LeCun, L.~Bottou, Y.~Bengio, and P.~Haffner, ``Gradient-based learning
  applied to document recognition,'' \emph{Proceedings of the IEEE}, vol.~86,
  no.~11, pp. 2278--2324, 1998.

\bibitem{xiao2017fashion}
H.~Xiao, K.~Rasul, and R.~Vollgraf, ``Fashion-mnist: a novel image dataset for
  benchmarking machine learning algorithms,'' \emph{arXiv preprint
  arXiv:1708.07747}, 2017.

\bibitem{krizhevsky2009learning}
A.~Krizhevsky, G.~Hinton \emph{et~al.}, ``Learning multiple layers of features
  from tiny images,'' 2009.

\bibitem{darlow2018cinic}
L.~N. Darlow, E.~J. Crowley, A.~Antoniou, and A.~J. Storkey, ``Cinic-10 is not
  imagenet or cifar-10,'' \emph{arXiv preprint arXiv:1810.03505}, 2018.

\bibitem{yang2019federated}
Q.~Yang, Y.~Liu, T.~Chen, and Y.~Tong, ``Federated machine learning: Concept
  and applications,'' \emph{ACM Transactions on Intelligent Systems and
  Technology (TIST)}, vol.~10, no.~2, pp. 1--19, 2019.

\bibitem{reddi2021adaptive}
S.~Reddi, Z.~Charles, M.~Zaheer, Z.~Garrett, K.~Rush, J.~Kone{\v{c}}n{\`y},
  S.~Kumar, and H.~B. McMahan, ``Adaptive federated optimization,''
  \emph{International Conference on Learning Representations}, 2021.

\bibitem{guha2019one}
N.~Guha, A.~Talwalkar, and V.~Smith, ``One-shot federated learning,''
  \emph{arXiv preprint arXiv:1902.11175}, 2019.

\bibitem{zhang2022dense}
\BIBentryALTinterwordspacing
J.~Zhang, C.~Chen, B.~Li, L.~Lyu, S.~Wu, S.~Ding, C.~Shen, and C.~Wu,
  ``{DENSE}: Data-free one-shot federated learning,'' in \emph{Advances in
  Neural Information Processing Systems}, A.~H. Oh, A.~Agarwal, D.~Belgrave,
  and K.~Cho, Eds., 2022. [Online]. Available:
  \url{https://openreview.net/forum?id=QFQoxCFYEkA}
\BIBentrySTDinterwordspacing

\bibitem{he2020group}
C.~He, M.~Annavaram, and S.~Avestimehr, ``Group knowledge transfer: Federated
  learning of large cnns at the edge,'' \emph{arXiv preprint arXiv:2007.14513},
  2020.

\bibitem{wu2022exploring}
Z.~Wu, S.~Sun, M.~Liu, J.~Zhang, Y.~Wang, and Q.~Liu, ``Exploring the
  distributed knowledge congruence in proxy-data-free federated distillation,''
  \emph{arXiv preprint arXiv:2204.07028}, 2022.

\bibitem{itahara2021distill}
S.~Itahara, T.~Nishio, Y.~Koda, M.~Morikura, and K.~Yamamoto,
  ``Distillation-based semi-supervised federated learning for
  communication-efficient collaborative training with non-iid private data,''
  \emph{IEEE Transactions on Mobile Computing}, pp. 1--1, 2021.

\bibitem{liu2020client}
L.~Liu, J.~Zhang, S.~Song, and K.~B. Letaief, ``Client-edge-cloud hierarchical
  federated learning,'' in \emph{ICC 2020-2020 IEEE International Conference on
  Communications (ICC)}.\hskip 1em plus 0.5em minus 0.4em\relax IEEE, 2020, pp.
  1--6.

\bibitem{wang2022accelerating}
Z.~Wang, H.~Xu, J.~Liu, Y.~Xu, H.~Huang, and Y.~Zhao, ``Accelerating federated
  learning with cluster construction and hierarchical aggregation,'' \emph{IEEE
  Transactions on Mobile Computing}, 2022.

\bibitem{xu2022hierfedml}
Z.~Xu, D.~Zhao, W.~Liang, O.~F. Rana, P.~Zhou, M.~Li, W.~Xu, H.~Li, and Q.~Xia,
  ``Hierfedml: aggregator placement and ue assignment for hierarchical
  federated learning in mobile edge computing,'' \emph{IEEE Transactions on
  Parallel and Distributed Systems}, vol.~34, no.~1, pp. 328--345, 2022.

\bibitem{tan2020federated}
B.~Tan, B.~Liu, V.~Zheng, and Q.~Yang, ``A federated recommender system for
  online services,'' in \emph{Fourteenth ACM Conference on Recommender
  Systems}, 2020, pp. 579--581.

\bibitem{zhou2022sourcetarget}
\BIBentryALTinterwordspacing
X.~Zhou, Y.~Tian, and X.~Wang, ``Source-target unified knowledge distillation
  for memory-efficient federated domain adaptation on edge devices,'' 2022.
  [Online]. Available: \url{https://openreview.net/forum?id=8rCMq0yJMG}
\BIBentrySTDinterwordspacing

\bibitem{jiang2020customized}
H.~Jiang, M.~Liu, B.~Yang, Q.~Liu, J.~Li, and X.~Guo, ``Customized federated
  learning for accelerated edge computing with heterogeneous task targets,''
  \emph{Computer Networks}, vol. 183, p. 107569, 2020.

\bibitem{wu2023survey}
Z.~Wu, S.~Sun, Y.~Wang, M.~Liu, X.~Jiang, and R.~Li, ``Survey of knowledge
  distillation in federated edge learning,'' \emph{arXiv preprint
  arXiv:2301.05849}, 2023.

\bibitem{zhang2022fedzkt}
L.~Zhang, D.~Wu, and X.~Yuan, ``Fedzkt: Zero-shot knowledge transfer towards
  resource-constrained federated learning with heterogeneous on-device
  models,'' in \emph{2022 IEEE 42nd International Conference on Distributed
  Computing Systems (ICDCS)}.\hskip 1em plus 0.5em minus 0.4em\relax IEEE,
  2022, pp. 928--938.

\bibitem{yu2022resource}
S.~Yu, W.~Qian, and A.~Jannesari, ``Resource-aware federated learning using
  knowledge extraction and multi-model fusion,'' \emph{arXiv preprint
  arXiv:2208.07978}, 2022.

\bibitem{arivazhagan2019federated}
M.~G. Arivazhagan, V.~Aggarwal, A.~K. Singh, and S.~Choudhary, ``Federated
  learning with personalization layers,'' \emph{arXiv preprint
  arXiv:1912.00818}, 2019.

\bibitem{liu2022communication}
L.~Liu, J.~Zhang, S.~Song, and K.~B. Letaief, ``Communication-efficient
  federated distillation with active data sampling,'' in \emph{ICC 2022-IEEE
  International Conference on Communications}.\hskip 1em plus 0.5em minus
  0.4em\relax IEEE, 2022, pp. 201--206.

\end{thebibliography}


\begin{thebibliography}{10}
\providecommand{\url}[1]{#1}
\csname url@samestyle\endcsname
\providecommand{\newblock}{\relax}
\providecommand{\bibinfo}[2]{#2}
\providecommand{\BIBentrySTDinterwordspacing}{\spaceskip=0pt\relax}
\providecommand{\BIBentryALTinterwordstretchfactor}{4}
\providecommand{\BIBentryALTinterwordspacing}{\spaceskip=\fontdimen2\font plus
\BIBentryALTinterwordstretchfactor\fontdimen3\font minus \fontdimen4\font\relax}
\providecommand{\BIBforeignlanguage}[2]{{%
\expandafter\ifx\csname l@#1\endcsname\relax
\typeout{** WARNING: IEEEtran.bst: No hyphenation pattern has been}%
\typeout{** loaded for the language `#1'. Using the pattern for}%
\typeout{** the default language instead.}%
\else
\language=\csname l@#1\endcsname
\fi
#2}}
\providecommand{\BIBdecl}{\relax}
\BIBdecl

\bibitem{ren2019survey}
J.~Ren, D.~Zhang, S.~He, Y.~Zhang, and T.~Li, ``A survey on end-edge-cloud orchestrated network computing paradigms: Transparent computing, mobile edge computing, fog computing, and cloudlet,'' \emph{ACM Computing Surveys (CSUR)}, vol.~52, no.~6, pp. 1--36, 2019.

\bibitem{duan2022distributed}
S.~Duan, D.~Wang, J.~Ren, F.~Lyu, Y.~Zhang, H.~Wu, and X.~Shen, ``Distributed artificial intelligence empowered by end-edge-cloud computing: A survey,'' \emph{IEEE Communications Surveys \& Tutorials}, vol.~25, no.~1, pp. 591--624, 2022.

\bibitem{wang2024end}
Y.~Wang, C.~Yang, S.~Lan, L.~Zhu, and Y.~Zhang, ``End-edge-cloud collaborative computing for deep learning: A comprehensive survey,'' \emph{IEEE Communications Surveys \& Tutorials}, 2024.

\bibitem{radford2019language}
A.~Radford, J.~Wu, R.~Child, D.~Luan, D.~Amodei, I.~Sutskever \emph{et~al.}, ``Language models are unsupervised multitask learners,'' \emph{OpenAI blog}, vol.~1, no.~8, p.~9, 2019.

\bibitem{vaswani2017attention}
A.~Vaswani, N.~Shazeer, N.~Parmar, J.~Uszkoreit, L.~Jones, A.~N. Gomez, {\L}.~Kaiser, and I.~Polosukhin, ``Attention is all you need,'' \emph{Advances in neural information processing systems}, vol.~30, 2017.

\bibitem{naveed2023comprehensive}
H.~Naveed, A.~U. Khan, S.~Qiu, M.~Saqib, S.~Anwar, M.~Usman, N.~Akhtar, N.~Barnes, and A.~Mian, ``A comprehensive overview of large language models,'' \emph{arXiv preprint arXiv:2307.06435}, 2023.

\bibitem{openai2022gpt3-5}
\BIBentryALTinterwordspacing
OpenAI, ``Gpt-3.5 turbo: Legacy gpt model for cheaper chat and non-chat tasks,'' 2022. [Online]. Available: \url{https://platform.openai.com/docs/models/gpt-3.5-turbo}
\BIBentrySTDinterwordspacing

\bibitem{brown2020language}
T.~Brown, B.~Mann, N.~Ryder, M.~Subbiah, J.~D. Kaplan, P.~Dhariwal, A.~Neelakantan, P.~Shyam, G.~Sastry, A.~Askell \emph{et~al.}, ``Language models are few-shot learners,'' \emph{Advances in neural information processing systems}, vol.~33, pp. 1877--1901, 2020.

\bibitem{guo2025deepseek}
D.~Guo, D.~Yang, H.~Zhang, J.~Song, R.~Zhang, R.~Xu, Q.~Zhu, S.~Ma, P.~Wang, X.~Bi \emph{et~al.}, ``Deepseek-r1: Incentivizing reasoning capability in llms via reinforcement learning,'' \emph{arXiv preprint arXiv:2501.12948}, 2025.

\bibitem{liu2024mobilellm}
Z.~Liu, C.~Zhao, F.~Iandola, C.~Lai, Y.~Tian, I.~Fedorov, Y.~Xiong, E.~Chang, Y.~Shi, R.~Krishnamoorthi \emph{et~al.}, ``Mobilellm: Optimizing sub-billion parameter language models for on-device use cases,'' in \emph{Forty-first International Conference on Machine Learning}, 2024.

\bibitem{hu2024minicpm}
S.~Hu, Y.~Tu, X.~Han, C.~He, G.~Cui, X.~Long, Z.~Zheng, Y.~Fang, Y.~Huang, W.~Zhao \emph{et~al.}, ``Minicpm: Unveiling the potential of small language models with scalable training strategies,'' \emph{arXiv preprint arXiv:2404.06395}, 2024.

\bibitem{fu2024serverlessllm}
Y.~Fu, L.~Xue, Y.~Huang, A.-O. Brabete, D.~Ustiugov, Y.~Patel, and L.~Mai, ``$\{$ServerlessLLM$\}$:$\{$Low-Latency$\}$ serverless inference for large language models,'' in \emph{18th USENIX Symposium on Operating Systems Design and Implementation (OSDI 24)}, 2024, pp. 135--153.

\bibitem{zhong2024distserve}
Y.~Zhong, S.~Liu, J.~Chen, J.~Hu, Y.~Zhu, X.~Liu, X.~Jin, and H.~Zhang, ``$\{$DistServe$\}$: Disaggregating prefill and decoding for goodput-optimized large language model serving,'' in \emph{18th USENIX Symposium on Operating Systems Design and Implementation (OSDI 24)}, 2024, pp. 193--210.

\bibitem{yang2024perllm}
Z.~Yang, Y.~Yang, C.~Zhao, Q.~Guo, W.~He, and W.~Ji, ``Perllm: Personalized inference scheduling with edge-cloud collaboration for diverse llm services,'' \emph{arXiv preprint arXiv:2405.14636}, 2024.

\bibitem{lin2023pushing}
Z.~Lin, G.~Qu, Q.~Chen, X.~Chen, Z.~Chen, and K.~Huang, ``Pushing large language models to the 6g edge: Vision, challenges, and opportunities,'' \emph{arXiv preprint arXiv:2309.16739}, 2023.

\bibitem{lebovitz2023efficient}
L.~Lebovitz, L.~Cavigelli, M.~Magno, and L.~K. Muller, ``Efficient inference with model cascades,'' \emph{Transactions on Machine Learning Research}, 2023.

\bibitem{yang2024qwen2}
A.~Yang, B.~Yang, B.~Zhang, B.~Hui, B.~Zheng, B.~Yu, C.~Li, D.~Liu, F.~Huang, H.~Wei \emph{et~al.}, ``Qwen2. 5 technical report,'' \emph{arXiv preprint arXiv:2412.15115}, 2024.

\bibitem{devlin2019bert}
J.~Devlin, M.-W. Chang, K.~Lee, and K.~Toutanova, ``Bert: Pre-training of deep bidirectional transformers for language understanding,'' in \emph{Proceedings of the 2019 conference of the North American chapter of the association for computational linguistics: human language technologies, volume 1 (long and short papers)}, 2019, pp. 4171--4186.

\bibitem{liu2019roberta}
Y.~Liu, M.~Ott, N.~Goyal, J.~Du, M.~Joshi, D.~Chen, O.~Levy, M.~Lewis, L.~Zettlemoyer, and V.~Stoyanov, ``Roberta: A robustly optimized bert pretraining approach,'' \emph{arXiv preprint arXiv:1907.11692}, 2019.

\bibitem{radford2018improving}
A.~Radford, K.~Narasimhan, T.~Salimans, I.~Sutskever \emph{et~al.}, ``Improving language understanding by generative pre-training,'' 2018.

\bibitem{hu2022lora}
E.~J. Hu, Y.~Shen, P.~Wallis, Z.~Allen-Zhu, Y.~Li, S.~Wang, L.~Wang, W.~Chen \emph{et~al.}, ``Lora: Low-rank adaptation of large language models.'' \emph{ICLR}, vol.~1, no.~2, p.~3, 2022.

\bibitem{houlsby2019parameter}
N.~Houlsby, A.~Giurgiu, S.~Jastrzebski, B.~Morrone, Q.~De~Laroussilhe, A.~Gesmundo, M.~Attariyan, and S.~Gelly, ``Parameter-efficient transfer learning for nlp,'' in \emph{International conference on machine learning}.\hskip 1em plus 0.5em minus 0.4em\relax PMLR, 2019, pp. 2790--2799.

\bibitem{ouyang2022training}
L.~Ouyang, J.~Wu, X.~Jiang, D.~Almeida, C.~Wainwright, P.~Mishkin, C.~Zhang, S.~Agarwal, K.~Slama, A.~Ray \emph{et~al.}, ``Training language models to follow instructions with human feedback,'' \emph{Advances in neural information processing systems}, vol.~35, pp. 27\,730--27\,744, 2022.

\bibitem{raffel2020exploring}
C.~Raffel, N.~Shazeer, A.~Roberts, K.~Lee, S.~Narang, M.~Matena, Y.~Zhou, W.~Li, and P.~J. Liu, ``Exploring the limits of transfer learning with a unified text-to-text transformer,'' \emph{Journal of machine learning research}, vol.~21, no. 140, pp. 1--67, 2020.

\bibitem{duan2023}
S.~Duan, D.~Wang, J.~Ren, F.~Lyu, Y.~Zhang, H.~Wu, and X.~Shen, ``Distributed artificial intelligence empowered by end-edge-cloud computing: A survey,'' \emph{IEEE Communications Surveys \& Tutorials}, vol.~25, no.~1, pp. 591--624, 2023.

\bibitem{yang2023hierarchical}
Z.~Yang, S.~Fu, W.~Bao, D.~Yuan, and A.~Y. Zomaya, ``Hierarchical federated learning with momentum acceleration in multi-tier networks,'' \emph{IEEE Transactions on Parallel and Distributed Systems}, 2023.

\bibitem{wu2025beyond}
Z.~Wu, S.~Sun, Y.~Wang, M.~Liu, K.~Xu, Q.~Pan, B.~Gao, and T.~Wen, ``Beyond model scale limits: End-edge-cloud federated learning with self-rectified knowledge agglomeration,'' \emph{arXiv preprint arXiv:2501.00693}, 2025.

\bibitem{shi2016edge}
W.~Shi, J.~Cao, Q.~Zhang, Y.~Li, and L.~Xu, ``Edge computing: Vision and challenges,'' \emph{IEEE internet of things journal}, vol.~3, no.~5, pp. 637--646, 2016.

\bibitem{wu2024agglomerative}
Z.~Wu, S.~Sun, Y.~Wang, M.~Liu, B.~Gao, Q.~Pan, T.~He, and X.~Jiang, ``Agglomerative federated learning: Empowering larger model training via end-edge-cloud collaboration,'' in \emph{IEEE INFOCOM 2024-IEEE Conference on Computer Communications}.\hskip 1em plus 0.5em minus 0.4em\relax IEEE, 2024, pp. 131--140.

\bibitem{islam2021survey}
A.~Islam, A.~Debnath, M.~Ghose, and S.~Chakraborty, ``A survey on task offloading in multi-access edge computing,'' \emph{Journal of Systems Architecture}, vol. 118, p. 102225, 2021.

\bibitem{saeik2021task}
F.~Saeik, M.~Avgeris, D.~Spatharakis, N.~Santi, D.~Dechouniotis, J.~Violos, A.~Leivadeas, N.~Athanasopoulos, N.~Mitton, and S.~Papavassiliou, ``Task offloading in edge and cloud computing: A survey on mathematical, artificial intelligence and control theory solutions,'' \emph{Computer Networks}, vol. 195, p. 108177, 2021.

\bibitem{dong2024task}
S.~Dong, J.~Tang, K.~Abbas, R.~Hou, J.~Kamruzzaman, L.~Rutkowski, and R.~Buyya, ``Task offloading strategies for mobile edge computing: A survey,'' \emph{Computer Networks}, p. 110791, 2024.

\bibitem{pearce2021understanding}
T.~Pearce, A.~Brintrup, and J.~Zhu, ``Understanding softmax confidence and uncertainty,'' \emph{arXiv preprint arXiv:2106.04972}, 2021.

\bibitem{maas2011learning}
A.~Maas, R.~E. Daly, P.~T. Pham, D.~Huang, A.~Y. Ng, and C.~Potts, ``Learning word vectors for sentiment analysis,'' in \emph{Proceedings of the 49th annual meeting of the association for computational linguistics: Human language technologies}, 2011, pp. 142--150.

\bibitem{socher2013recursive}
R.~Socher, A.~Perelygin, J.~Wu, J.~Chuang, C.~D. Manning, A.~Y. Ng, and C.~Potts, ``Recursive deep models for semantic compositionality over a sentiment treebank,'' in \emph{Proceedings of the 2013 conference on empirical methods in natural language processing}, 2013, pp. 1631--1642.

\bibitem{pang2005seeing}
B.~Pang and L.~Lee, ``Seeing stars: Exploiting class relationships for sentiment categorization with respect to rating scales,'' \emph{arXiv preprint cs/0506075}, 2005.

\bibitem{zhang2015character}
X.~Zhang, J.~Zhao, and Y.~LeCun, ``Character-level convolutional networks for text classification,'' \emph{Advances in neural information processing systems}, vol.~28, 2015.

\bibitem{bojar2016findings}
O.~Bojar, R.~Chatterjee, C.~Federmann, Y.~Graham, B.~Haddow, M.~Huck, A.~J. Yepes, P.~Koehn, V.~Logacheva, C.~Monz \emph{et~al.}, ``Findings of the 2016 conference on machine translation (wmt16),'' in \emph{First conference on machine translation}.\hskip 1em plus 0.5em minus 0.4em\relax Association for Computational Linguistics, 2016, pp. 131--198.

\bibitem{barrault2019findings}
L.~Barrault, O.~Bojar, M.~R. Costa-Jussa, C.~Federmann, M.~Fishel, Y.~Graham, B.~Haddow, M.~Huck, P.~Koehn, S.~Malmasi \emph{et~al.}, ``Findings of the 2019 conference on machine translation (wmt19).''\hskip 1em plus 0.5em minus 0.4em\relax ACL, 2019.

\bibitem{tiedemann2012parallel}
J.~Tiedemann, ``Parallel data, tools and interfaces in opus.'' in \emph{Lrec}, vol. 2012.\hskip 1em plus 0.5em minus 0.4em\relax Citeseer, 2012, pp. 2214--2218.

\bibitem{Sanh2019DistilBERTAD}
V.~Sanh, L.~Debut, J.~Chaumond, and T.~Wolf, ``Distilbert, a distilled version of bert: smaller, faster, cheaper and lighter,'' \emph{ArXiv}, vol. abs/1910.01108, 2019.

\bibitem{seq2class-end}
\BIBentryALTinterwordspacing
azizbarank, ``azizbarank/distilroberta-base-sst2-distilled,'' 2022. [Online]. Available: \url{https://huggingface.co/azizbarank/distilroberta-base-sst2-distilled}
\BIBentrySTDinterwordspacing

\bibitem{seq2class-edge}
\BIBentryALTinterwordspacing
textattack, ``textattack/roberta-base-sst-2,'' 2023. [Online]. Available: \url{https://huggingface.co/textattack/roberta-base-SST-2}
\BIBentrySTDinterwordspacing

\bibitem{seq2class-cloud}
\BIBentryALTinterwordspacing
howey, ``howey/roberta-large-sst2,'' 2021. [Online]. Available: \url{https://huggingface.co/howey/roberta-large-sst2}
\BIBentrySTDinterwordspacing

\bibitem{seq2seq-end}
\BIBentryALTinterwordspacing
SEBIS, ``Sebis/legal_t5_small_trans_de_en_small_finetuned,'' 2021. [Online]. Available: \url{https://huggingface.co/SEBIS/legal_t5_small_trans_de_en_small_finetuned}
\BIBentrySTDinterwordspacing

\bibitem{tiedemann2020opus}
J.~Tiedemann and S.~Thottingal, ``Opus-mt--building open translation services for the world,'' in \emph{Annual Conference of the European Association for Machine Translation}.\hskip 1em plus 0.5em minus 0.4em\relax European Association for Machine Translation, 2020, pp. 479--480.

\bibitem{seq2seq-edge}
\BIBentryALTinterwordspacing
PontifexMaximus, ``Pontifexmaximus/opus-mt-de-en-finetuned-de-to-en,'' 2022. [Online]. Available: \url{https://huggingface.co/PontifexMaximus/opus-mt-de-en-finetuned-de-to-en}
\BIBentrySTDinterwordspacing

\bibitem{kasai2020deep}
J.~Kasai, N.~Pappas, H.~Peng, J.~Cross, and N.~A. Smith, ``Deep encoder, shallow decoder: Reevaluating non-autoregressive machine translation,'' \emph{arXiv preprint arXiv:2006.10369}, 2020.

\bibitem{seq2seq-cloud}
\BIBentryALTinterwordspacing
allenai, ``allenai/wmt19-de-en-6-6-big,'' 2023. [Online]. Available: \url{https://huggingface.co/allenai/wmt19-de-en-6-6-big}
\BIBentrySTDinterwordspacing

\bibitem{padmanabhan2023gemel}
A.~Padmanabhan, N.~Agarwal, A.~Iyer, G.~Ananthanarayanan, Y.~Shu, N.~Karianakis, G.~H. Xu, and R.~Netravali, ``Gemel: Model merging for $\{$Memory-Efficient$\}$,$\{$Real-Time$\}$ video analytics at the edge,'' in \emph{20th USENIX Symposium on Networked Systems Design and Implementation (NSDI 23)}, 2023, pp. 973--994.

\bibitem{he2020deberta}
P.~He, X.~Liu, J.~Gao, and W.~Chen, ``Deberta: Decoding-enhanced bert with disentangled attention,'' \emph{arXiv preprint arXiv:2006.03654}, 2020.

\bibitem{ablation-model-deberta}
\BIBentryALTinterwordspacing
Tomor0720, ``Tomor0720/deberta-large-finetuned-sst2,'' 2023. [Online]. Available: \url{https://huggingface.co/Tomor0720/deberta-large-finetuned-sst2}
\BIBentrySTDinterwordspacing

\bibitem{achiam2023gpt}
J.~Achiam, S.~Adler, S.~Agarwal, L.~Ahmad, I.~Akkaya, F.~L. Aleman, D.~Almeida, J.~Altenschmidt, S.~Altman, S.~Anadkat \emph{et~al.}, ``Gpt-4 technical report,'' \emph{arXiv preprint arXiv:2303.08774}, 2023.

\bibitem{shahriar2024putting}
S.~Shahriar, B.~D. Lund, N.~R. Mannuru, M.~A. Arshad, K.~Hayawi, R.~V.~K. Bevara, A.~Mannuru, and L.~Batool, ``Putting gpt-4o to the sword: A comprehensive evaluation of language, vision, speech, and multimodal proficiency,'' \emph{Applied Sciences}, vol.~14, no.~17, p. 7782, 2024.

\bibitem{claude3}
\BIBentryALTinterwordspacing
{Anthropic}, ``{The Claude 3 Model Family: Opus, Sonnet, Haiku},'' 2024. [Online]. Available: \url{https://www-cdn.anthropic.com/de8ba9b01c9ab7cbabf5c33b80b7bbc618857627/Model\_Card\_Claude\_3.pdf}
\BIBentrySTDinterwordspacing

\bibitem{claude3-5}
\BIBentryALTinterwordspacing
Anthropic, ``Claude 3-5 sonnet,'' 2024. [Online]. Available: \url{https://www.anthropic.com/news/claude-3-5-sonnet}
\BIBentrySTDinterwordspacing

\bibitem{claude3-7}
\BIBentryALTinterwordspacing
{Anthropic}, ``Claude 3.7 sonnet and claude code,'' 2025. [Online]. Available: \url{https://www.anthropic.com/news/claude-3-7-sonnet}
\BIBentrySTDinterwordspacing

\bibitem{touvron2023llama}
H.~Touvron, T.~Lavril, G.~Izacard, X.~Martinet, M.-A. Lachaux, T.~Lacroix, B.~Rozi{\`e}re, N.~Goyal, E.~Hambro, F.~Azhar \emph{et~al.}, ``Llama: Open and efficient foundation language models,'' \emph{arXiv preprint arXiv:2302.13971}, 2023.

\bibitem{touvron2023llama2}
H.~Touvron, L.~Martin, K.~Stone, P.~Albert, A.~Almahairi, Y.~Babaei, N.~Bashlykov, S.~Batra, P.~Bhargava, S.~Bhosale \emph{et~al.}, ``Llama 2: Open foundation and fine-tuned chat models,'' \emph{arXiv preprint arXiv:2307.09288}, 2023.

\bibitem{grattafiori2024llama}
A.~Grattafiori, A.~Dubey, A.~Jauhri, A.~Pandey, A.~Kadian, A.~Al-Dahle, A.~Letman, A.~Mathur, A.~Schelten, A.~Vaughan \emph{et~al.}, ``The llama 3 herd of models,'' \emph{arXiv preprint arXiv:2407.21783}, 2024.

\bibitem{bai2023qwen}
J.~Bai, S.~Bai, Y.~Chu, Z.~Cui, K.~Dang, X.~Deng, Y.~Fan, W.~Ge, Y.~Han, F.~Huang \emph{et~al.}, ``Qwen technical report,'' \emph{arXiv preprint arXiv:2309.16609}, 2023.

\bibitem{wang2024qwen2}
P.~Wang, S.~Bai, S.~Tan, S.~Wang, Z.~Fan, J.~Bai, K.~Chen, X.~Liu, J.~Wang, W.~Ge \emph{et~al.}, ``Qwen2-vl: Enhancing vision-language model's perception of the world at any resolution,'' \emph{arXiv preprint arXiv:2409.12191}, 2024.

\bibitem{bai2025qwen2}
S.~Bai, K.~Chen, X.~Liu, J.~Wang, W.~Ge, S.~Song, K.~Dang, P.~Wang, S.~Wang, J.~Tang \emph{et~al.}, ``Qwen2. 5-vl technical report,'' \emph{arXiv preprint arXiv:2502.13923}, 2025.

\bibitem{kwon2023efficient}
W.~Kwon, Z.~Li, S.~Zhuang, Y.~Sheng, L.~Zheng, C.~H. Yu, J.~Gonzalez, H.~Zhang, and I.~Stoica, ``Efficient memory management for large language model serving with pagedattention,'' in \emph{Proceedings of the 29th Symposium on Operating Systems Principles}, 2023, pp. 611--626.

\bibitem{liu2024cachegen}
Y.~Liu, H.~Li, Y.~Cheng, S.~Ray, Y.~Huang, Q.~Zhang, K.~Du, J.~Yao, S.~Lu, G.~Ananthanarayanan \emph{et~al.}, ``Cachegen: Kv cache compression and streaming for fast large language model serving,'' in \emph{Proceedings of the ACM SIGCOMM 2024 Conference}, 2024, pp. 38--56.

\bibitem{shubha2023adainf}
S.~S. Shubha and H.~Shen, ``Adainf: Data drift adaptive scheduling for accurate and slo-guaranteed multiple-model inference serving at edge servers,'' in \emph{Proceedings of the ACM SIGCOMM 2023 Conference}, 2023, pp. 473--485.

\bibitem{zhang2024edgeshard}
M.~Zhang, X.~Shen, J.~Cao, Z.~Cui, and S.~Jiang, ``Edgeshard: Efficient llm inference via collaborative edge computing,'' \emph{IEEE Internet of Things Journal}, 2024.

\bibitem{matsubara2022split}
Y.~Matsubara, M.~Levorato, and F.~Restuccia, ``Split computing and early exiting for deep learning applications: Survey and research challenges,'' \emph{ACM Computing Surveys}, vol.~55, no.~5, pp. 1--30, 2022.

\bibitem{mohammed2020distributed}
T.~Mohammed, C.~Joe-Wong, R.~Babbar, and M.~Di~Francesco, ``Distributed inference acceleration with adaptive dnn partitioning and offloading,'' in \emph{IEEE INFOCOM 2020-IEEE conference on computer communications}.\hskip 1em plus 0.5em minus 0.4em\relax IEEE, 2020, pp. 854--863.

\bibitem{xu2022mandheling}
D.~Xu, M.~Xu, Q.~Wang, S.~Wang, Y.~Ma, K.~Huang, G.~Huang, X.~Jin, and X.~Liu, ``Mandheling: Mixed-precision on-device dnn training with dsp offloading,'' in \emph{Proceedings of the 28th Annual International Conference on Mobile Computing And Networking}, 2022, pp. 214--227.

\bibitem{pan2024instinfer}
X.~Pan, E.~Li, Q.~Li, S.~Liang, Y.~Shan, K.~Zhou, Y.~Luo, X.~Wang, and J.~Zhang, ``Instinfer: In-storage attention offloading for cost-effective long-context llm inference,'' \emph{arXiv preprint arXiv:2409.04992}, 2024.

\end{thebibliography}

\vspace{-33pt}

 \begin{IEEEbiography}[{\includegraphics[width=1in,height=1.25in,clip,keepaspectratio]{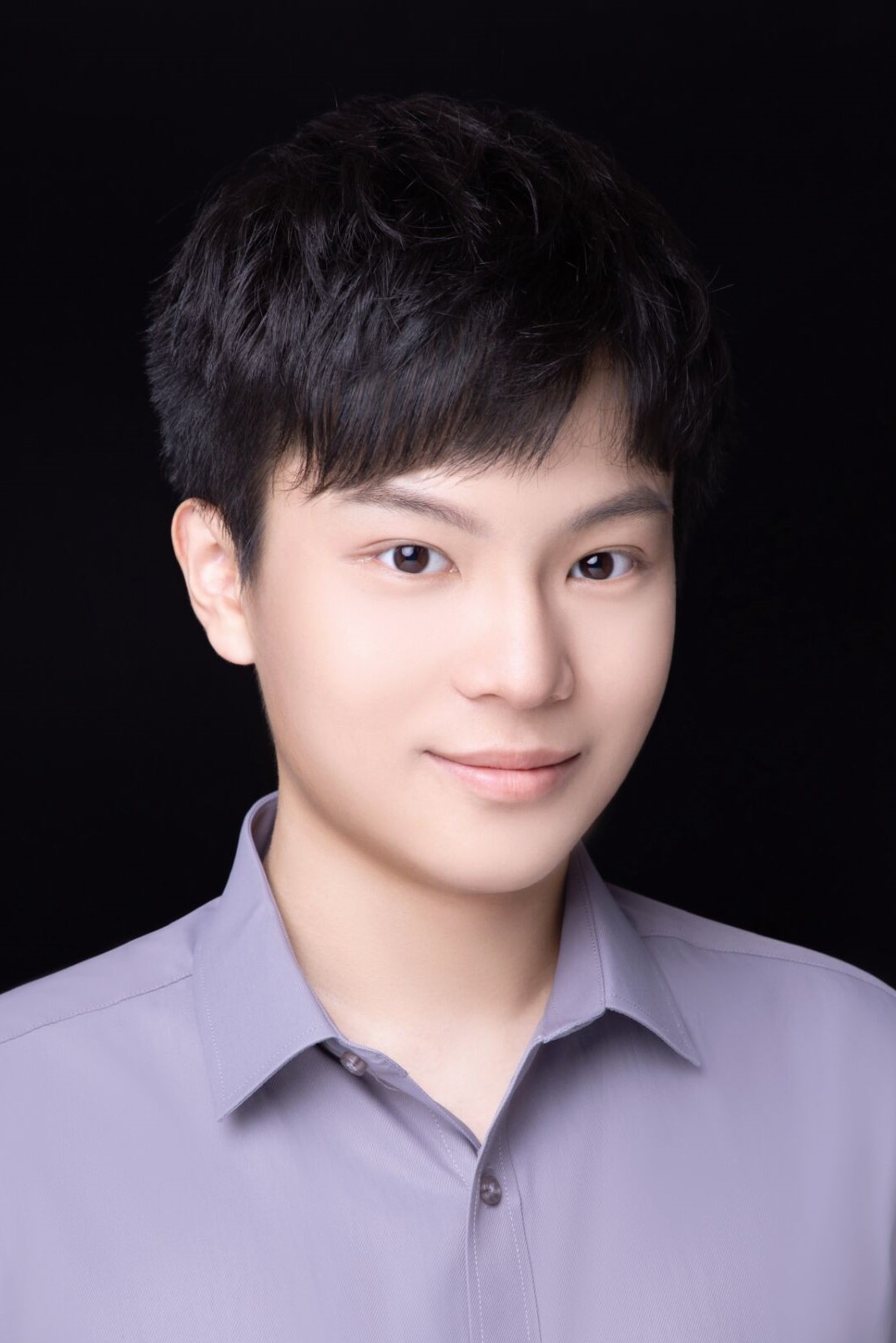}}]{Zhiyuan Wu} is currently a research assistant with the Institute of Computing Technology, Chinese Academy of Sciences. He has contributed several technical papers to top-tier conferences and journals as the first author in the fields of computer architecture, computer networks, and intelligent systems, including IEEE Transactions on Parallel and Distributed Systems (TPDS), IEEE Transactions on Mobile Computing (TMC), IEEE International Conference on Computer Communications (INFOCOM), and ACM Transactions on Intelligent Systems and Technology (TIST). His research interests include edge intelligence, distributed systems, and services computing.
\end{IEEEbiography}
 
 \begin{IEEEbiography}[{\includegraphics[width=1in,height=1.25in,clip,keepaspectratio]{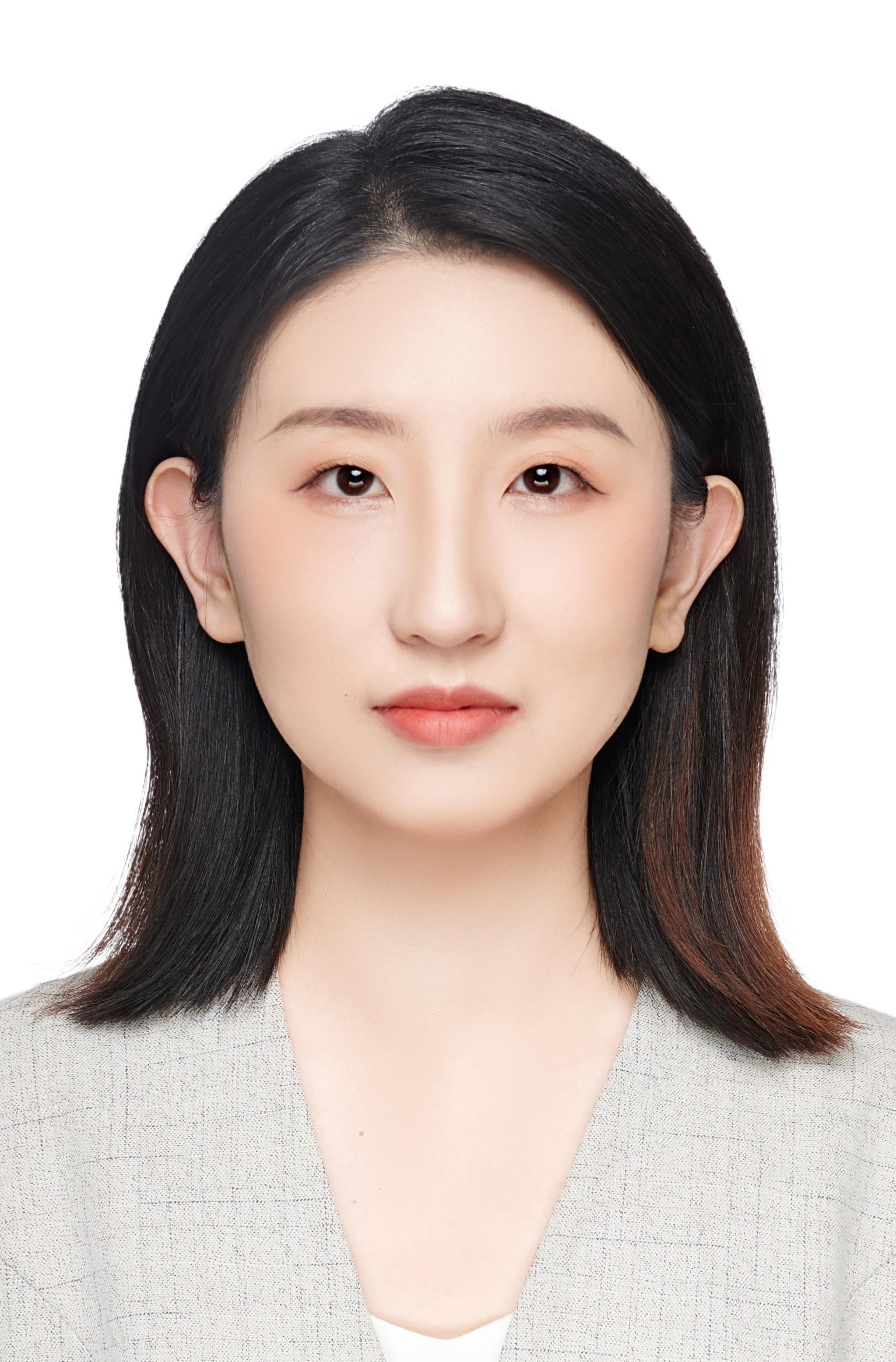}}]{Sheng Sun} is currently an associate professor at the Institute of Computing Technology, Chinese Academy of Sciences. She received her bachelor's degree from Beihang University, and her Ph.D. from the Institute of Computing Technology, Chinese Academy of Sciences. Dr. Sun has led or executed 5 major funded research projects and published over 20 technical papers in journals and conferences related to computer network and distributed systems, including IEEE Transactions on Parallel and Distributed Systems (TPDS), IEEE Transactions on Mobile Computing (TMC), and IEEE International Conference on Computer Communications (INFOCOM). Her research interests include federated learning, edge intelligence, and privacy computing.
\end{IEEEbiography}

\begin{IEEEbiography}[{\includegraphics[width=1in,height=1.25in,clip,keepaspectratio]{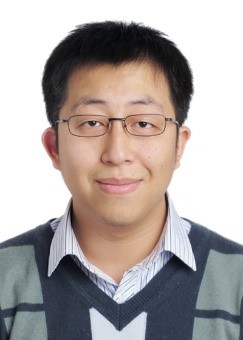}}]{Yuwei Wang} (Member, IEEE) received his Ph.D. degree in computer science from the University of Chinese Academy of Sciences, Beijing, China. He is currently an associate professor at the Institute of Computing Technology, Chinese Academy of Sciences. He has been responsible for setting over 30 international and national standards, and also holds various positions in both international and national industrial standards development organizations (SDOs) as well as local research institutions, including the associate rapporteur at the ITU-T SG21 Q5, and the deputy director of China Communications Standards Association (CCSA) TC1 WG1. His current research interests include federated learning, mobile edge computing, and next-generation network architecture.
\end{IEEEbiography}

\begin{IEEEbiography}[{\includegraphics[width=1in,height=1.25in,clip,keepaspectratio]{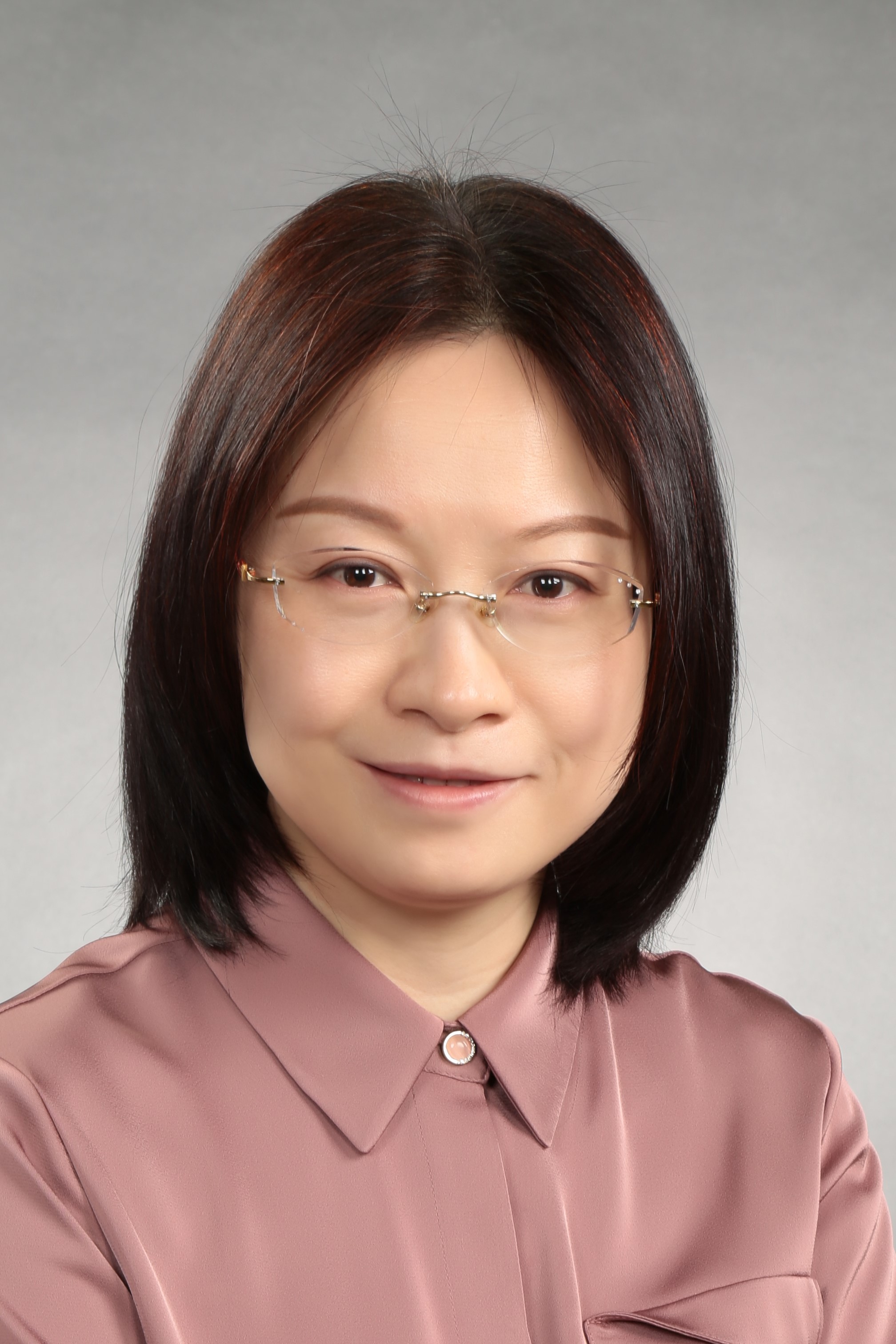}}]{Min Liu} (Senior Member, IEEE) received her Ph.D degree in computer science from the Graduate University of the Chinese Academy of Sciences, China. Before that, she received her B.S. and M.S. degrees in computer science from Xi’an Jiaotong University, China. She is currently a professor at the Institute of Computing Technology, Chinese Academy of Sciences, and also holds a position at the Zhongguancun Laboratory. Her current research interests include mobile computing and edge intelligence.
\end{IEEEbiography}

 \begin{IEEEbiography}[{\includegraphics[width=1in,height=1.25in,clip,keepaspectratio]{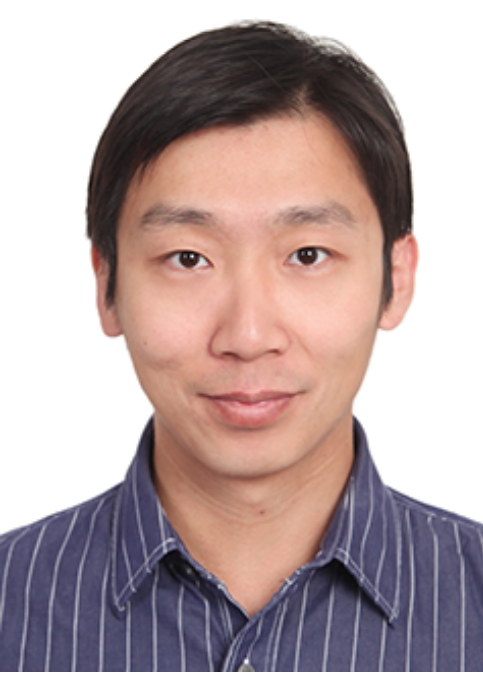}}]{Bo Gao} (Member, IEEE) received his M.S. degree in electrical engineering from the School of Electronic Information and Electrical Engineering at Shanghai Jiaotong University, Shanghai, China in 2009, and his Ph.D. degree in computer engineering from the Bradley Department of Electrical and Computer Engineering at Virginia Tech, Blacksburg, USA in 2014. He was an Assistant Professor with the Institute of Computing Technology at Chinese Academy of Sciences, Beijing, China from 2014 to 2017. He was a Visiting Researcher with the School of Computing and Communications at Lancaster University, Lancaster, UK from 2018 to 2019. He is currently an Associate Professor with the School of Computer and Information Technology at Beijing Jiaotong University, Beijing, China. He has directed a number of research projects sponsored by the National Natural Science Foundation of China (NSFC) or other funding agencies. He is a member of IEEE, ACM, and China Computer Federation (CCF). His research interests include wireless networking, mobile/edge computing, multiagent systems, and machine learning.
 \end{IEEEbiography}

\begin{IEEEbiography}[{\includegraphics[width=1in,height=1.25in,clip,keepaspectratio]{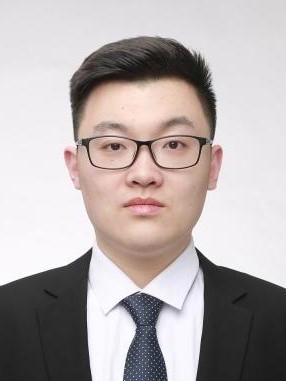}}]{Jinda Lu}
	is currently a Phd candidate with the School of Cyber Science and Technology, University of Science and Technology of China. Before that, he received his bachelor's degree with honor at Jilin University. His research interests include large language models and multi modality.
\end{IEEEbiography}

\begin{IEEEbiography}[{\includegraphics[width=1in,height=1.25in,clip,keepaspectratio]{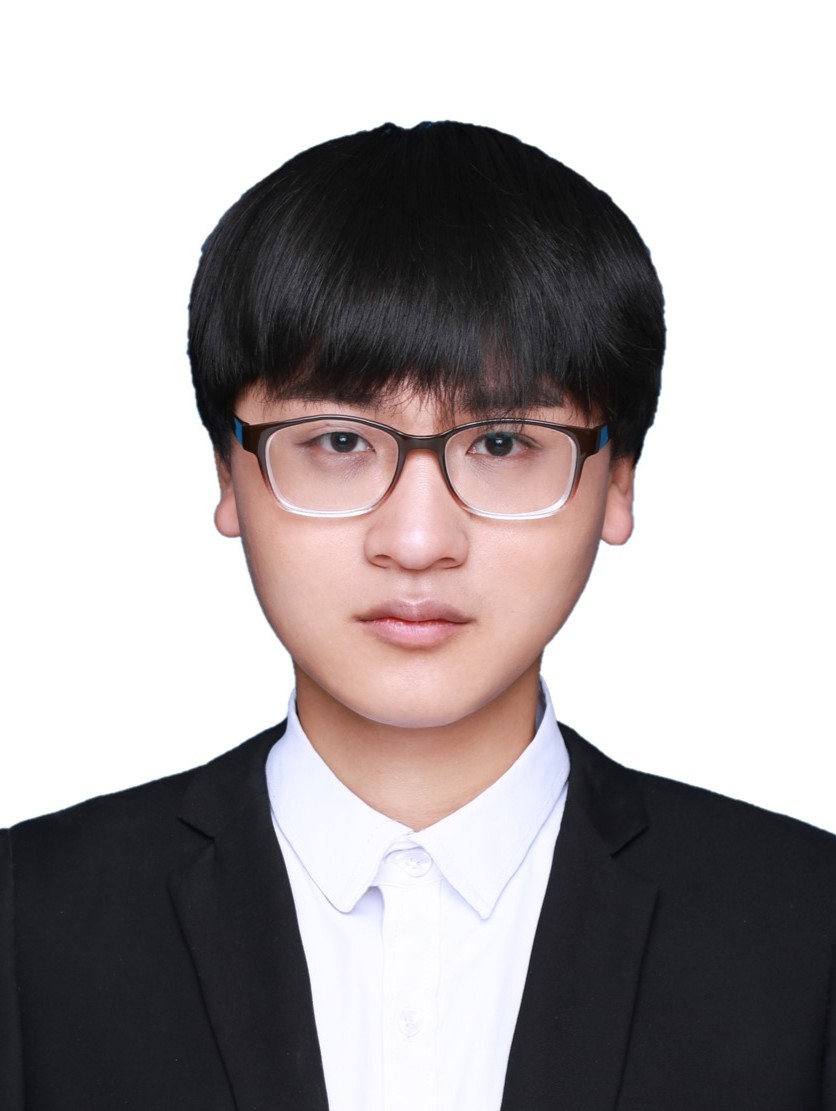}}]{Zheming Yang} received his B.Sc. degree in electronic engineering from the North China University of Science and Technology in 2019 and his Ph.D. degree in computer science from the Institute of Computing Technology, Chinese Academy of Sciences, in 2024. He
is currently an assistant professor with the Institute of Computing Technology, Chinese Academy of Sciences, Beijing, China. He is also a visiting scholar with the School of Computing, National University of Singapore. His current research interests include multimedia systems, edge/cloud computing, LLM inference optimization, and efficient AI system design. In addition, his research won the IFTC Best Paper Award, the IEEE ISPA Outstanding Paper Award, and the CCF HPC China Outstanding Paper Award.
\end{IEEEbiography}

\begin{IEEEbiography}[{\includegraphics[width=1in,height=1.25in,clip,keepaspectratio]{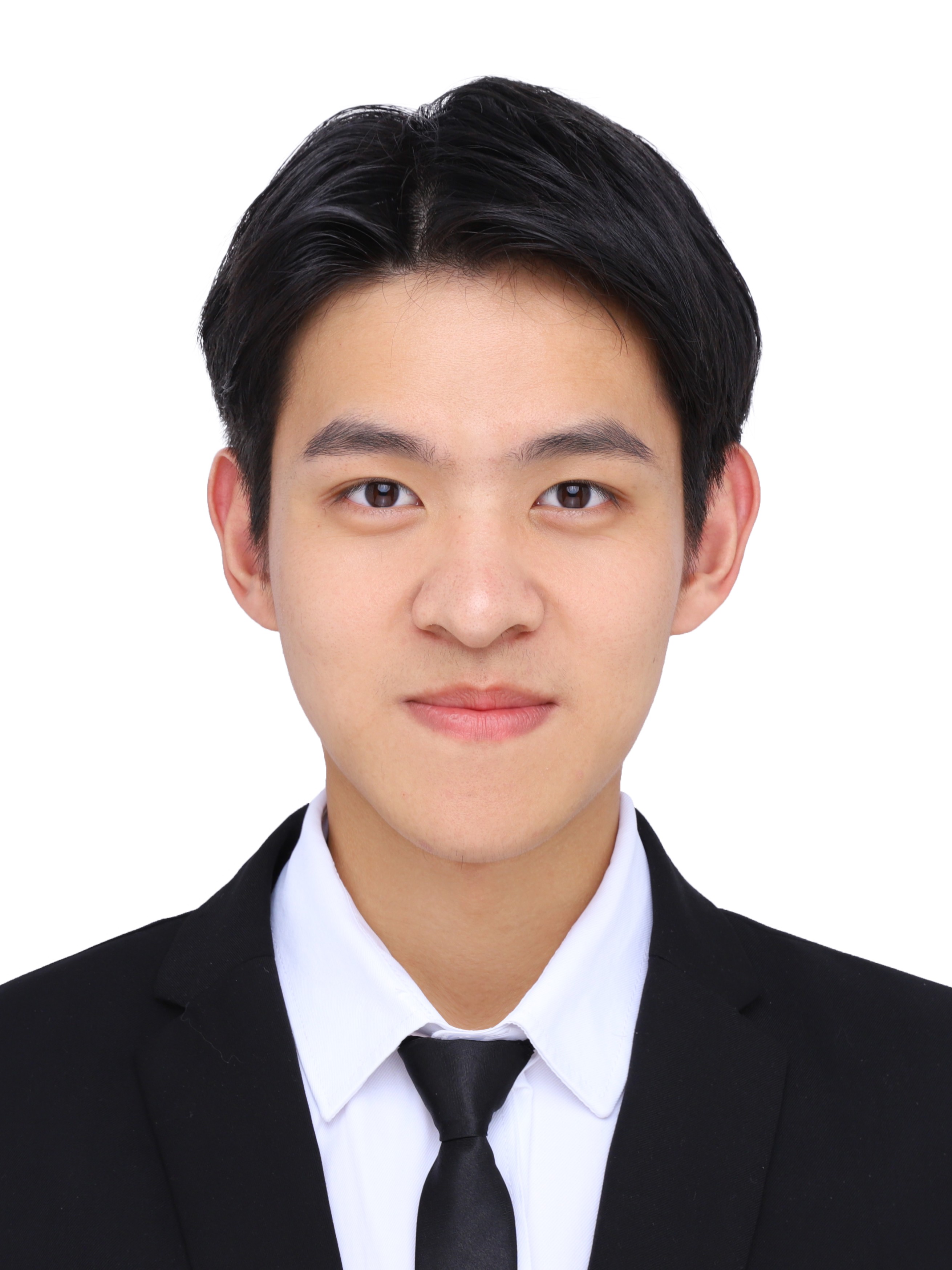}}]{Tian Wen} is currently a research assistant with the Institute of Computing Technology, Chinese Academy of Sciences. His research interests include federated learning, edge computing, and information security.
\end{IEEEbiography}

\vfill
\newpage
\appendices
\onecolumn
\end{document}